\documentclass[sigconf]{acmart}
\usepackage{endnotes,verbatim,booktabs,multirow,comment,tabularx,xcolor}

\newcommand{\blu}{}
\newcommand{\bk}{\textbf}

\copyrightyear{2024}
\acmYear{2024}
\setcopyright{acmlicensed}\acmConference[CHI '24]{Proceedings of the CHI Conference on Human Factors in Computing Systems}{May 11--16, 2024}{Honolulu, HI, USA}
\acmBooktitle{Proceedings of the CHI Conference on Human Factors in Computing Systems (CHI '24), May 11--16, 2024, Honolulu, HI, USA}
\acmDOI{10.1145/3613904.3642107}
\acmISBN{979-8-4007-0330-0/24/05}

\begin{document}

\title{Analysis and Implementation of Nanotargeting on LinkedIn Based on Publicly Available Non-PII}

\author{Ángel Merino}
\affiliation{%
    \institution{Universidad Carlos III de Madrid}
    \streetaddress{Av. Universidad 30}
    \city{Leganés}
    \country{Spain}
}
\email{angel.merino@uc3m.es}

\author{José González-Cabañas}
\affiliation{%
    \institution{UC3M-Santander Big Data Institute}
    \streetaddress{C. Madrid, 135}
    \city{Getafe}
    \country{Spain}
}
\email{jose.gonzalez.cabanas@uc3m.es}

\author{Ángel Cuevas}
\affiliation{%
    \institution{Universidad Carlos III de Madrid}
    \streetaddress{Av. Universidad 30}
    \city{Leganés}
    \country{Spain}
}
\email{acrumin@it.uc3m.es}

\author{Rubén Cuevas}
\affiliation{%
    \institution{Universidad Carlos III de Madrid}
    \streetaddress{Av. Universidad 30}
    \city{Leganés}
    \country{Spain}
}
\email{rcuevas@it.uc3m.es}
\additionalaffiliation{%
    \institution{UC3M-Santander Big Data Institute}
    \streetaddress{C. Madrid, 135}
    \city{Getafe}
    \country{Spain}
}

\begin{abstract}
The literature has shown that combining a few non-Personal Identifiable Information (non-PII) is enough to make a user unique in a dataset including millions of users. This work demonstrates that a combination of a few non-PII items can be activated to nanotarget users. We demonstrate that the combination of the location and {5} rare ({13} random) skills in a LinkedIn profile is enough to become unique in a user base of {$\sim$970M} users with a probability of 75\%. The novelty is that these attributes are publicly accessible to anyone registered on LinkedIn and can be activated through advertising campaigns. We ran an experiment configuring ad campaigns using the location and skills of three of the paper's authors, demonstrating how all the ads using $\geq13$ skills were delivered exclusively to the targeted user. We reported this vulnerability to LinkedIn, which initially ignored the problem, but fixed it as of November 2023.%This nanotargeting may expose LinkedIn users to privacy and security risks such as malvertising or manipulation.
\end{abstract}




%\received{14 September 2023}
%\received[revised]{}
%\received[accepted]{}

\maketitle
\section{Introduction}
\label{sec:introduction}

Current data protection regulations such as the GDPR \cite{gdpr} clearly state that PII is personal data and requires (in most cases) the user's consent to be processed. However, the combination of multiple non-PII items that are not considered personal data in isolation may make a user unique, thus constituting personal data. This is why user uniqueness based on non-PII data has been addressed in the literature in recent years. For example, just 4 mobile phone call records can identify a user in a dataset of 1.5 million users \cite{de2013unique}. Similarly, in a user base of 1.1 million users, only 4 credit card purchase records are needed to single out an individual \cite{de2015unique}. Also, 15 demographic attributes can re-identify 99.98\% of Americans in any dataset \cite{Nature_unique}. However, all those works remain theoretical and do not discuss how the non-PII data items can be activated in specific attacks compromising users' security and/or privacy. The natural step forward to complete this area of research is developing methodologies and experiments to demonstrate that the combination of non-PII items can be activated in practice by third parties to target users individually and (potentially) compromise their security and/or privacy.

{Despite there were prior research works \cite{prank, harf_2017,  tim_shipman_2018, hawkins_2019, faddoul2019sniper, haskins_2018,faizullabhoy2018facebooks,venkatadriPIIfb} showing that PII data can be used to individually target an individual with ads}, to the best of the authors' knowledge, the only prior study in the field that practically shows that a combination of non-PII items can be activated to \textit{reach} a single user exclusively with an ad is \cite{fbnano}. In this work we performed a proof of concept experiment, showing that an attacker being able to unveil $\sim$20  random ad preferences from a user in Facebook can target them with a \emph{nanotargeting} ad campaign, i.e., the ad reaches the targeted user exclusively. This is the first tangible proof that non-PII information can be exploited to target individual users without explicit consent to be reached uniquely by those means. However, the practical use of the reported technique at scale has a significant limitation. It requires the attacker to access users' ad preferences, which is a complex task since they are not publicly available. This limitation reduces the potential attackers to those with strong technical knowledge able to infer the ad preferences of a user. Although the referred work is a very relevant research contribution, we believe it is important that the research community contributes further studies showing that it is feasible to implement hyper-personalized attacks subject to publicly available non-PII items proactively disclosed by users. Such studies would prove that non-PII items, often not considered personal data, may involve severe privacy and/or security risks for users.

Our work shows that hundreds of millions of users could have been individually targeted with hyper-tailored ads combining a few non-PII publicly available data items before the LinkedIn fix. To this end, in this research, we imposed ourselves three requirements: $(i)$ the user base should include tens or hundreds of millions of users distributed all over the world; $(ii)$ the non-PII data items required to target an individual user must be publicly available, and $(iii)$ the non-PII items can be activated by external third-parties to reach users with hyper-personalized ads individually. To the best of our knowledge, none of the previous works in the literature meet these three requirements simultaneously.

Our paper proves that an individual user could have been nanotargeted on LinkedIn with an ad using the combination of the location (country, region, or city) and the professional skills available in their profile. This meets the three previous requirements as follows: $(i)$ LinkedIn has $\sim$1000 million users, i.e., roughly 12\% of the worldwide population, $(ii)$ the location and professional skills of LinkedIn users are publicly available non-PII items to anyone logged on LinkedIn. Hence, anyone can easily obtain the required information that uniquely identifies a user on LinkedIn, and ($iii$) the combination of professional skills and location can be activated through the LinkedIn Ads Manager to deliver hyper-personalized ads to the users. In practice, this means that nanotargeting a user just required having a LinkedIn account, retrieving the location and professional skills from the targeted user profile, and configuring an ad campaign using that information. This was a very simple operation that may have enabled many third parties willing to do so to run nanotargeting campaigns/attacks on LinkedIn exploiting non-PII items before November 2023. Therefore, we considerably extend the contribution of our previous work \cite{fbnano} because we demonstrate for the first time that nanotargeting could be implemented using publicly available information. This implies a dramatic increase in the risk users may have been exposed to because many more attackers could implement nanotargeting attacks.

{The psychological persuasion literature has demonstrated that using tailored ads (e.g., ad creativity) to an individual substantially increases the probability of persuading and/or attracting the attention of the targeted individual \cite{pshyc_persuasion1}\cite{pshyc_persuasion2}\cite{pshyc_persuasion3}\cite{pshyc_persuasion4}\cite{pshyc_persuasion5}\cite{Kosinski_PNAS}. Nanotargeting may allow malicious advertisers running attacks such as user manipulation \cite{youtube} or malvertising \cite{Malvertising_Bayesian}\cite{Malvertising_SVM} \cite{Malvertising_mobile1}\cite{Malvertising_NDSS}\cite{Malvertising_WWW} \cite{Malvertising_Aritz}. Therefore, the threat model associated with nanotargeting is that an advertiser may act as an attacker through malvertising attacks to perform phishing, spread malware, etc, or manipulation attacks to persuade users to carry out actions aligned to the attacker's interests \cite{youtube}.} 

We divided our work into two parts. In the first part of the paper, we use a dataset including information on 78k skills collected from 3352 users, and we develop a data-driven model that defines the probability of user uniqueness on LinkedIn by combining the location and N professional skills publicly available in their profile. Based on this result, we compute a lower bound estimation suggesting that a quarter billion LinkedIn users may have been exposed to nanotargeting attacks with a 95\% probability of success. In the second part of the paper, we use the model's outcome to implement a proof of concept experiment, targeting three authors of this paper, demonstrating that it was feasible to run nanotargeting campaigns on LinkedIn before November 2023. 

LinkedIn's ad guidelines state a minimum of 300 targeted members for a campaign, but we could bypass this limit by exploiting an implementation bug. We reported the privacy vulnerability unveiled by our research to LinkedIn in April 2023, following their recommended process. Unfortunately, the platform managers considered our findings did not constitute a vulnerability and closed the issue.

We kept working on our research and run nanotargeting campaigns until July 2023. Surprisingly, while working on revising our manuscript out of the review process of ACM CHI 2024, in November 2023, we realized LinkedIn had fixed the problem and running nanotargeting campaigns was not possible anymore. We are very happy LinkedIn fixed the vulnerability because this was the ultimate purpose of our research and because this improves the privacy and security of hundreds of millions of users. %Although someone could think this was unfortunate to us because it reduced the relevance of our research, we do not share that point of view.
In our view, this confirms the relevance of our findings. 

We believe it is extremely important to publish success stories around academic works in the area of privacy to demonstrate that it is feasible to achieve practical positive impacts through research. In our opinion, the ultimate goal of any privacy researcher should be to have a real impact on improving the privacy/security of citizens. However, it is unfortunate LinkedIn did not give us any credit for this, but we are equally happy because our goal has been achieved.

To conclude the introduction we summarize the key findings of our work:

\begin{itemize}
    \item Combining users' location with 13 (21) randomly selected skills from their reported skill set makes them unique on LinkedIn with a 75\% (90\%) probability. If we use the least popular skills instead, we only need 5 (14) skills to achieve the same level of uniqueness. 
    \item Our proof of concept experiment shows that all campaigns using the location and $\geq$13 random skills successfully nanotargeted the three targeted authors.
    \item To the best of our knowledge, this is the first study showing proof that publicly available non-PII data could have been used to effectively target unique citizens \emph{at scale}, before LinkedIn fixed the bug.
    \item This work represents an example of how privacy researchers can have a practical impact and improve the privacy of citizens.
    %In other words, anyone registered on LinkedIn could potentially send exclusive ads to a single user based on the user's non-PII items publicly shared on their profile.
\end{itemize}

\section{LinkedIn background and nanotargeting threat model}
\label{sec:background}

In this section, we introduce the LinkedIn advertising platform, discuss the differences between using LinkedIn messages and ads to target users, and why we are focusing on the latter, and describe the threat model and risks associated with nanotargeting advertising campaigns.

\subsection{LinkedIn advertising platform}

One of LinkedIn's primary sources of revenue is online advertising \cite{ldrevenue}. Advertisers use the LinkedIn Campaign Manager \cite{ldcampaignmanager} to define their target audience by selecting attributes such as location, gender, age, professional skills, etc. Then LinkedIn delivers the ads to users whose profile matches the selected attributes. For example, an advertiser could target \emph{users in Germany between 18 and 25 who are skilled in Python and R}.

LinkedIn reports the estimated number of users that match the target audience defined through the dashboard, referred to as \emph{Target Audience Size}. This information permits advertisers to know the potential reach of their ad campaign. In our work, we leverage this functionality and use specific HTTP queries to systematically collect the size of thousands of audiences.

An essential feature of the LinkedIn Campaign Manager is that it allows us to narrow the audience size by combining non-PII attributes using the AND operator. Moreover, the information of users (location, skills, experience, education, etc.) is publicly available in their profile for anyone registered on LinkedIn \cite{ldvisibility}. We have explored whether it is possible to customize the access to a user profile (e.g., to limit the access only to direct contacts), and it was not possible. It is important to note that LinkedIn is a professional social network where users are willing to expose their profiles publicly. This may be why LinkedIn does not consider offering users profile visibility customization options relevant.
  
LinkedIn seems to consider that revealing the actual value of small audiences may be problematic and does not report the actual audience size for any audience formed by less than 300 users. This is a positive privacy-preserving measure, a standard among social media platforms with a similar business model. For instance, Facebook and TikTok have defined a lower bound of 1000 users. Moreover, we found a LinkedIn document including guidelines for advertisers where they specifically state: \textit{"The minimum audience size required to run an advertising campaign is 300 members"} \cite{ldhelpaudience}. We understand this as a policy imposed by LinkedIn on ad campaigns. This measure would impede advertisers (attackers) from running nanotargeting campaigns to target a specific individual. Unfortunately, this paper demonstrates that this policy was not effectively enforced before November 2023, since it was possible to nanotarget LinkedIn users.

Finally, this paper focuses on the use of professional skills as non-PII items. The maximum number of skills a user can include in their profile is 50, and LinkedIn reports that the number of different skills available in their platform is +41k \cite{ldstats}.

\subsection{LinkedIn messages vs. ads}

LinkedIn already allows reaching individual users through the message functionality. When you visit the profile of a user on LinkedIn you may find a blue button with the text "Message". If you click it in one of your LinkedIn network connections a pop-up window appears where you can message that person. However, this has some limitations. First, standard LinkedIn users cannot use the messaging to reach LinkedIn members who are not 1st level contacts. The only option for that is to become a premium member. Still, message sending is limited. Once the quota is used up, payment is required for each additional message. Therefore, using messaging to reach non-contact users at scale is costly. Additionally, users may deactivate the messaging function to avoid being reached. For instance, Bill Gates (when writing this paper) does not have the message functionality enabled. We have analyzed 120 profiles from the LinkedIn Top Voices 2020 \cite{ldtopvoices} and $\sim$10\% of them have the messaging feature disabled. Finally, most users are familiar with interpreting individual messages from various platforms like email, SMS, and social media, so they are skilled at assessing the risks associated with messages from unknown senders. 

In the case of ads, most LinkedIn users are unaware that they can be targeted based on skills, making the portion of users disabling this option negligible, although it is possible. Therefore, any single user on LinkedIn is exposed to receiving ads. Furthermore, this paper analyzes how a positive feature such as showcasing skills for sharing professional expertise may be exploited to deliver hyper-personalized ads. Running nanotargeting ad campaigns was almost free, as we show in Section \ref{sec:poc}. Finally, users are less trained to distinguish between ads that may represent a threat and benign ads. 

However, the main reason for prioritizing ads is the legal contrast between messaging and nanotargeting ad campaigns. When you register on LinkedIn you are consenting to receive direct messages (unless you deactivate them). However, nowhere in the Terms of Use of LinkedIn or the privacy policy you are consenting to receive individually targeted ads based on non-PII data. Even more, LinkedIn's policies allow direct messaging but they state that it is not possible to deliver ads to audiences lower than 300 users. Therefore, if we consider modern data protection regulations such as the GDPR \cite{gdpr} direct messaging is perfectly compliant based on valid user consent. Contrarily, nanotargeting ad campaigns likely breach regulations, posing a clear threat to user privacy. 

Overall, we acknowledge the messaging feature is an interesting option for delivering individual messages, and it would be worth analyzing the efficiency and the cost of running nanotargeting attacks through it. However, this paper focuses on nanotargeting using ads because it may represent a non-compliant practice according to modern data protection regulations.

\subsection{Nanotargeting threat model}

As we mentioned in the introduction, the psychological persuasion literature has demonstrated that using hyper-personalized ads to target an individual increases the likelihood of persuading them and/or attracting their attention. Nanotargeting ad campaigns are a very powerful tool to deliver hyper-personalized messages embedded in ad creativities and increase the possibilities of the user engaging with the ad and the messages it includes. There are multiple risks users could be exposed to in case an attacker nanotarget them with tailored ads. Next, we briefly discuss a few of them.

-\noindent \textit{Malvertising}: Malvertising stands for malicious advertising and refers to the process of using online advertising to perform some attack \cite{Malvertising_Bayesian,Malvertising_SVM,Malvertising_mobile1,Malvertising_NDSS,Malvertising_WWW,Malvertising_Aritz}. Multiple vector attacks can be implemented through malvertising. One type of malvertising consists of using online ads to inject malware into ad networks, webpages, or the end-user device \cite{flashmalvert,zarrasmalvert,maladscollection,Mansfield-Devine-malvert}. A second type of malvertising replicates the concept of phishing attacks, but instead of using emails to capture the user's attention, it uses ads. This attack aims to persuade the user to click on the ad to land on a website managed by the attacker. At this point, the attacker, as in the case of phishing attacks, can use any potential technique willing to obtain sensitive information from the user (e.g., credentials) or to compromise the user's device. Nanotargeting may be especially relevant in this second type of attack emulating phishing. Nanotargeting allows the creation of hyper-personalized ads targeting a single user. The literature on psychological persuasion has demonstrated that a well-designed, tailored ad substantially increases the probability that the user clicks on the ad \cite{Kosinski_PNAS,mullock}. Since nanotargeting allows reaching the maximum expression of personalization, a savvy attacker can exploit it to increase the chances of persuading the users to click on the tailored ad and land in the attacker's domain. 

-\noindent \textit{User manipulation:} Our results demonstrated that we could run multiple campaigns over time to nanotarget an individual. In addition, our proof of concept experiment suggests that it was feasible to reach the users multiple times in a nanotargeting campaign in case they were active on LinkedIn. In other words, it was possible to expose the user to a tailored message frequently. Based on the psychological persuasion literature, it is easier to persuade an individual if the advertiser creates tailored ads to the psychological characteristics and motivations of that person \cite{pshyc_persuasion1,pshyc_persuasion2,pshyc_persuasion3,pshyc_persuasion4,pshyc_persuasion5,Kosinski_PNAS}. We also show a clear example of potential user influence in the related work \cite{youtube}. In that case, nanotargeting, based on PII data, was used to expose celebrities to a specific brand before approaching them to propose a collaboration. This same strategy could have been used to influence celebrities, CEOs, politicians, etc., with an account on LinkedIn. 

It is important to note that LinkedIn avoids malvertising attacks requiring the insertion of code in the ad since the LinkedIn ads platform does not allow this option. However, avoiding phishing-like malvertising or manipulation ad campaigns is a tough task since, if they are properly implemented, they do not differ from other regular advertising campaigns.

\section{Dataset}
\label{sec:dataset}

\begin{figure}[t]
    \includegraphics[width=1\columnwidth]{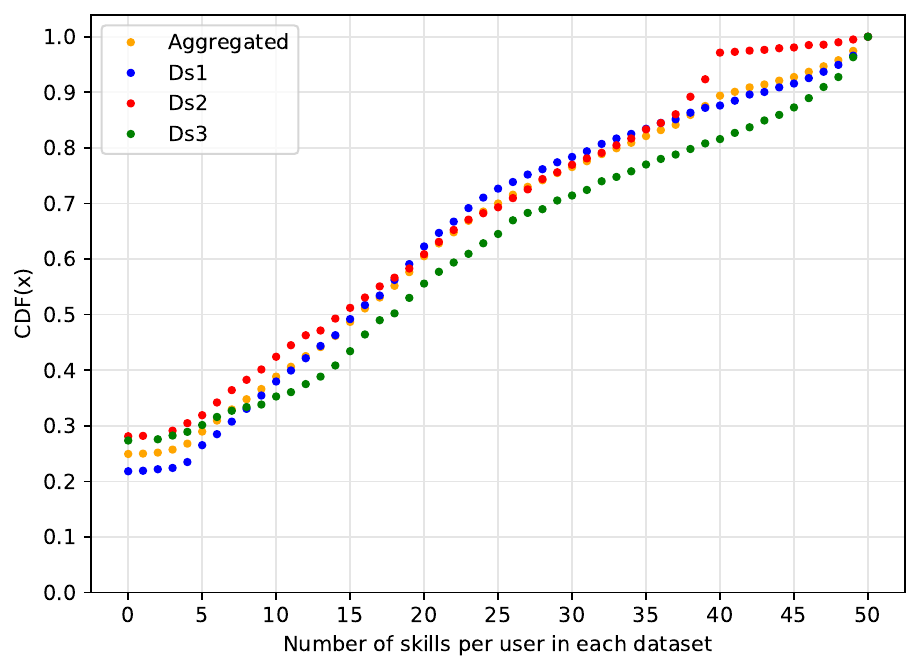}\hfill
    \caption{ CDF of the number of skills per user profile for our three data samples and the aggregated dataset.}
    \label{fig:CDF_skills_per_user}
    \Description{Dot chart showing the cumulative distribution function (CDF) of the number of skills per user in our dataset. The X-axis represents the number of skills, ranging from 1 to 50, and the Y axis represents the cumulative probability.}
\end{figure}

One of the contributions of this work is assessing whether combining location and professional skills can make a user unique on LinkedIn, {which is a compulsory requirement to implement nanotargeting attacks. To do that, we created a dataset including thousands of audience sizes for different combinations of locations and skills from real LinkedIn profiles to build a model (see Section \ref{sec:methodology}) that computes the number of skills required to make a user unique in LinkedIn with a probability $P$. The outcome of the model applied to our datasets (see Section \ref{sec:results})  will be later used to implement a proof of concept experiment to validate the feasibility of running nanotargeting attacks in LinkedIn  (see Section \ref{sec:poc}).}

We had to implement two different pieces of software to obtain the dataset. The first one retrieved the location and professional skills from thousands of LinkedIn users relying on different sampling methods. The second used the information from those profiles to retrieve the audience size from the LinkedIn Campaign Manager for thousands of audiences combining location and professional skills.

\begin{figure}[t]
    \includegraphics[width=1\columnwidth]{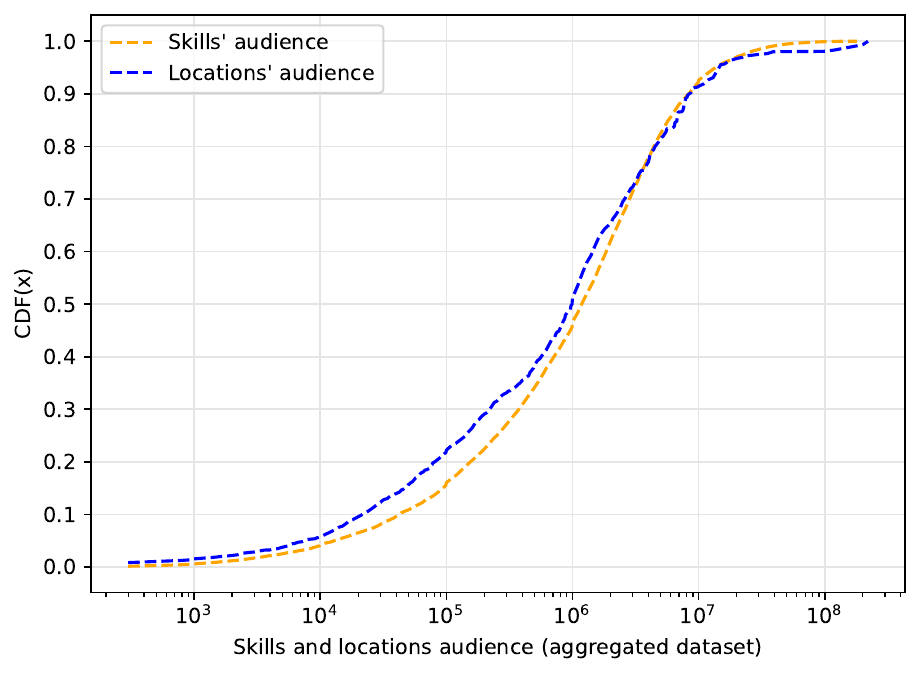}\hfill
    \caption{CDF of the worldwide audience size associated with the 8533 unique professional skills (orange line) and with the locations in our aggregated dataset (blue line).}
    \label{fig:cdf_popularity_skills}
    \Description{Cumulative distribution function (CDF) of the worldwide audience size associated with the 8533 unique professional skills (orange line) and with the locations (blue line) in our dataset. The X-axis represents the audience size values, and the Y-axis represents the cumulative probability.}
\end{figure}

\subsection{Users' skills and location}

We would have liked to collect a representative (i.e., random) dataset of LinkedIn profiles. Unfortunately, it is not feasible to formally verify whether a subsample of LinkedIn profiles is a random sample of the whole LinkedIn user base in the context of professional skills. That would require getting access to the actual distribution of professional skills on LinkedIn to verify the randomness of the collected subsample. To the best of our knowledge, this is not feasible for a third party different from LinkedIn.   

 To minimize this inherent limitation, we propose using three different sampling methods that aim to randomize the collection of LinkedIn profiles and analyze whether they report similar values for the number of skills that make a user unique with a probability P. If the three datasets lead to similar results, we would increase the confidence that the obtained outcome from the proposed model is a good reference to explain the uniqueness of LinkedIn users based on the number of skills they publicly report on their profiles.

The first dataset, referred to as Dataset 1 (Ds1), was created by searching LinkedIn profiles on DuckDuckGo. To that end, we used the Campaign Manager's job type classification. This field is known as \emph{job functions} and includes a general list of 26 professional fields. (i) We launched a search query on DuckDuckGo for each of the 26 job functions, using a filter to only obtain LinkedIn profiles, and (ii) retrieved the first ten results for each category. This led to 260 seed users from whom we collected the skills and locations they reported in their profiles. Starting with those users, we conducted a Breadth-First Search (BFS) \cite{BFS} and gathered information from the users LinkedIn suggested when accessing the seed profiles. Overall, we collected 2174 users from which 1690 have reported both skills and location, and 9 users have reported skills but not location. This means, 21.8\% of the profiles in the Ds1 did not include skills.

The second dataset, referred to as Dataset 2 (Ds2), was built directly by querying LinkedIn's search bar. (i) We selected 100 different countries and for each of them, we created one query per letter in the alphabet. This means, 2600 queries in total. Each query leads to 100 pages including profile entries. (ii) For each query, we got the URLs included in one of the pages selected at random. The purpose of this step is to differentiate Ds2 from Ds1 ensuring the selected profiles were not selecting the first entries returned in the search result, as it happened in Ds1. Overall, we had a pool of 19648 URLs pointing to LinkedIn profiles. (iii) From that pool, we randomly accessed 1396 profiles. 984 profiles included both skills and locations, 18 included skills but not location, and 394 (28\%) did not include any skill,  

To create the third dataset, referred to as Dataset 3 (Ds3), we (i) selected 764 skills from the user skills available in Ds1, (ii) launched a search in DuckDuckGo for each of those skills limiting the results to only LinkedIn profiles, and (iii) got the first 400 results (LinkedIn profiles). Again, this sampling method aims to avoid retrieving data from the first entries reported by LinkedIn and/or DuckDuckGo. Overall we created a pool of 328193 URLs to LinkedIn profiles, from which we crawled the information reported in 897 profiles. 643 of those profiles reported skills and locations, 8 skills but not location, and 246 (27\%) did not report any skill.

The aggregated dataset (DsAgg) includes 4467 users from which 25\% do not report any skill and 74\% report both skills and locations. From our sample of 3352 users reporting at least one skill, 3317 (99.96\%) also provided a location (country, region, or city). Overall, our dataset includes 8533 unique skills that appeared 78794 times across 3352 user profiles. This means that each unique skill was reported by 9 users in our dataset on average.

Figure \ref{fig:CDF_skills_per_user} shows the CDF of the number of skills reported per user for Ds1, Ds2, Ds3, and DsAgg. The median number of skills reported by users ranges between 14 (Ds2) and 18 skills (Ds3), with 15 skills if we consider the aggregated dataset. The results also show that around 30\% of the users include more than 25 skills in their profiles. The relevant portion of users not reporting skills (20\%-30\%) guarantees that our sampling methods are not biased toward users who report a large number of skills. 

\begin{figure}[t]
    \includegraphics[width=1\columnwidth]{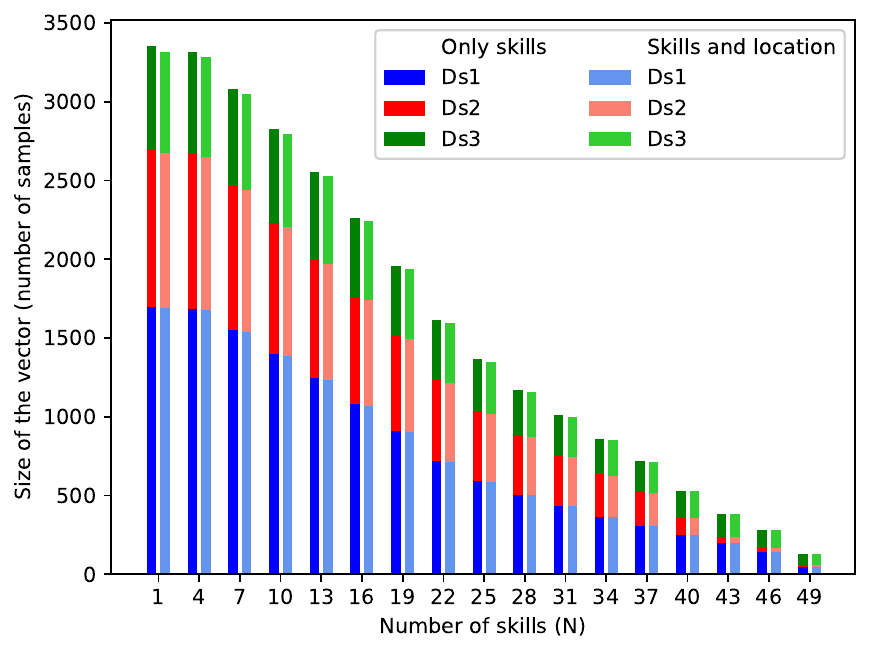}\hfill
    \caption{Length of the vectors used in our methodology according to the number of professional skills considered ranging from $N$=1 to $N$=50 skills. We show in different colors the portion of samples that corresponds to each dataset.}
    \label{fig:users_as_N_grows}
    \Description{Bar chart representing the number of samples with at least N skills listed. The X-axis represents the number of skills listed, and the Y axis represents the number of profiles that have listed at least that number of skills. For each value in the X axis, we represent two bars, one refers to the whole dataset, and the other only to the profiles that have a location listed.We show in different colors the portion of samples that corresponds to each dataset.}
\end{figure}

It is important to note that, from now on in the paper, we will only consider the profiles in our dataset that reported at least one skill. Following this decision, Figure \ref{fig:users_as_N_grows} shows the number of users in our dataset reporting between 1 and 50 skills (the maximum number of skills LinkedIn allows in a profile) in steps of 3.

Finally, we would like to note that we would have liked to create larger user samples in each dataset. However, the process of visiting each profile and getting their data is a bottleneck due to LinkedIn's limitations to profile visiting.

\color{black}
\subsection{Audience size}

{ Our methodology to infer the uniqueness probability of LinkedIn users does not work with profiles, but with the audience size associated with thousands of combinations of skills and location}. Therefore, we implemented ad hoc software to systematically obtain the size of LinkedIn audiences based on the skills and location of the 3352 individuals in our aggregated dataset. The data from those audience sizes feed the methodology outlined in the next section, which computes how many skills are necessary to make a user unique on LinkedIn.

{
Figure \ref{fig:cdf_popularity_skills} shows the CDF of the worldwide audience size associated with the 8533 unique professional skills in our final dataset. It also shows the audience size CDF of the locations associated with the 3352 individuals included in our dataset. A LinkedIn skill is featured in 1.2M profiles on the median, whereas the median audience size of the locations in our dataset is 1M.}

\subsection{Breakdown of the dataset users by country}

Our dataset contains samples from  107 \color{black} different countries, but a few countries constitute most of our dataset, especially the United States. Table \ref{table:samples_by_country} (in appendix \ref{app:countries_breakdown}) shows the breakdown of the number of users per country and dataset. About 49\% of the users in our dataset are from the United States. We acknowledge the fact that this circumstance in our data may lead to some biases in the results of our model and, therefore, to the estimation of $N$. However, the fact that the proof of concept experiment was targeting users in a different country than the US and the obtained results are aligned with the model outcome makes us confident that the potential bias (if any) may not be very relevant.  Also, Ds2 is not biased toward the United States, and the separate analysis of this dataset leads to similar results. \color{black}

Our intuition, based on the paper's results, is that our model serves as an upper bound for the number of skills needed to make a user unique. This may be because the United States is one of the countries contributing more LinkedIn users. Therefore, it seems reasonable to estimate that, in many cases, it will be easier to re-identify users reporting a different location than the US.

\section{Methodology}
\label{sec:methodology}

\begin{figure}[t]
\includegraphics[width=\hsize]{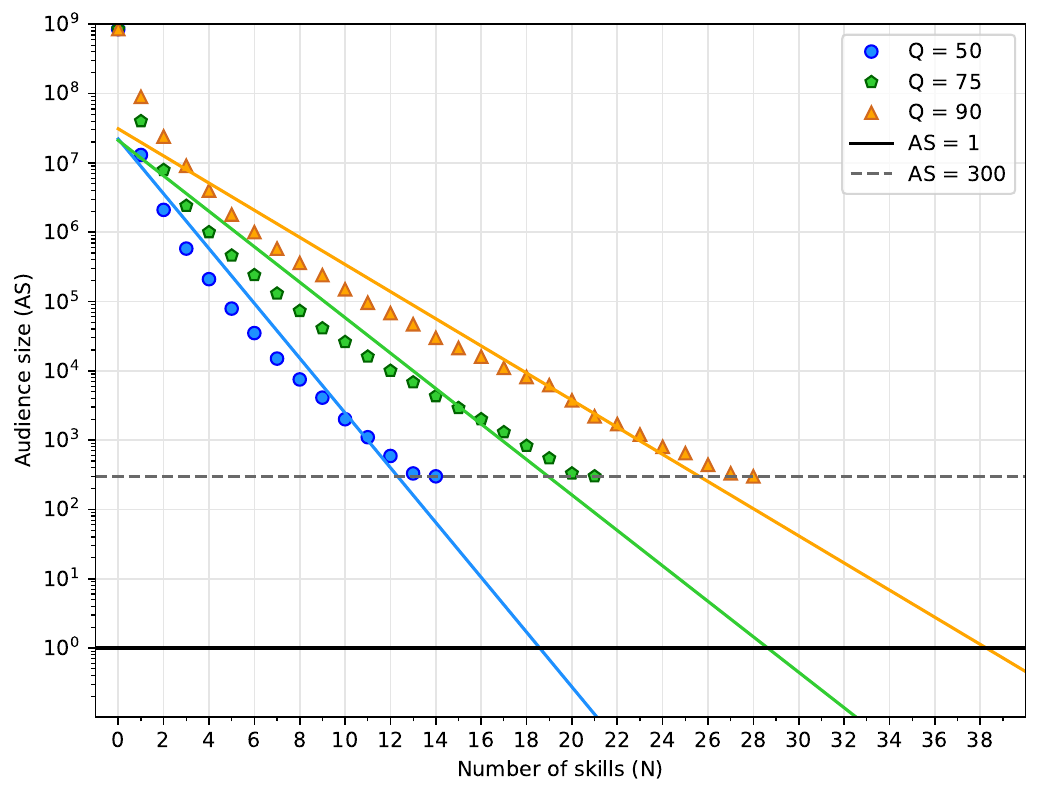}\hfill
	\caption{Application of the methodology to the Sk\_R\_Agg \color{black} scenario for $V\textsubscript{AS}(Q)$ with $Q = 50, 75$ and $90$. The figure visually depicts the model fitting (lines) to the data obtained from our dataset (markers). It also shows the audience size asymptote in 300 and a bold line where the audience size has a value equal to 1.}
	\label{fig:sk_r_line_fit}
    \Description{Line graph showing the results of applying our methodology to our aggregated dataset, for the scenario in which we only use skills to define the audiences (and not the location), and we select the skills randomly (Sk_R_Agg). The Y axis represents the audience size, while the X axis represents the number of skills used to define an audience. The points show the quantiles 50, 75, and 90 for each value in the X-axis, and the lines represent the fitting of the points for each quantile. Both the points and lines show how the audience size decreases as the audience is defined with more skills, and the lines estimate at what point the audience size would reach a value of one for each quantile. The estimations are approximately, 18 and 28 and 38 skills for the 50, 75, and 90 quantiles, respectively.}
\end{figure}

The first objective of this paper is to compute how many skills are required to uniquely identify a user on LinkedIn.  We rely on the methodology proposed in \cite{fbnano} and adapt it to the case of Linkedin. The outcome of this methodology is a metric referred to as $N\textsubscript{P}$. In the context of our work, this metric defines the number of skills $N$ that uniquely identify a user with probability $P$. For instance, $N\textsubscript{0.7} = 15$ means that knowing 15 skills from a user makes them unique with a probability of 70\%.

\subsection{Location dimension}
Using the location may substantially reduce the number of skills that make a user unique on LinkedIn. Because  99.96\% \color{black} of the profiles in our data sample released their location, it does not make sense not to use the location in a nanotargeting campaign in case it is available. Still, some users do not report their location. We aim to analyze both cases and estimate the number of skills that make a user unique when they report their location and when they do not report their location.

The initial conditions for the audience size in each of these cases differ. For instance, if we only consider skills, the starting point is the worldwide audience size reported by the LinkedIn Ads Manager when carrying out our research, which included  970M \color{black} users. In contrast, if we also consider the skills, the starting point for a given user is the audience size of the reported location. For instance, at the time of writing this paper, the audience size for the US, the state of New York, and New York City was: {\blu 220}M,  {\blu 14}M, and {\blu 8.2}M, respectively. { Intuitively, a user reporting New York City as the location will become unique with fewer skills than a user reporting the US.}

\subsection{Methodology to compute NP} 

For each user, $u\textsubscript{i} (i \in [1,{ 3352}])$, we leverage the LinkedIn Campaign Manager to obtain the audience size from a combination of $N$ skills with $N$ ranging from 1 to 50. We limit the number of skills to 50 because it is the highest number of skills a user can report on LinkedIn. 

For N=1, we obtain a vector with  3352 \color{black} different audience sizes by selecting one skill per user from the skills they have reported in their profile. For N=2, we get another vector with  3352 \color{black} different audience sizes by selecting two skills per user from the skills they have reported in their profile. We repeat the same operation for N = 3, 4, ..., 50.  At the end of the process, we have 50 vectors that show the audience size distribution for each value of $N$.

We note the number of users reporting $N$ skills in their profile is lower as $N$ increases.
Therefore for a given value of $N$, the vector will include as many samples as the number of users in our dataset that have reported $N$ or more skills in their profile. Figure \ref{fig:users_as_N_grows} shows the vector length for each value of $N$ and the two cases considered in our analysis: skills and location+skills.% We acknowledge that the number of samples for $N\geq 30$ may be small ($<$ 500 users), but as we will show in the next section, our methodology will not rely on those vectors to compute the number of skills that make a user unique on LinkedIn. 

Based on the constructed distributions, we define $AS(Q,N)$ as the audience size for the quantile $Q$ and the number of skills $N$. For instance, $AS(75,7) = 750$ means that the 75th percentile of an audience defined with 7 skills is 750 users. In other words, the probability that the audience size is $\leq 750$ using 7 skills is 75\%. We define $V_{AS}(Q)$ as a vector that includes all the 50 values of $AS(Q,N)$ for a fixed value of $Q$, i.e., $V_{AS}(Q) = [AS(Q,1), ..., AS(Q,50)]$. $V_{AS}(Q)$ is a decreasing vector by definition since $AS(Q,N) \geq AS(Q,N+1)$. At some point, we expect to find a value of $N$ for which $V_{AS}(Q)$ equals 1. That cutpoint is the output metric of the methodology $N_P$, where $P$ is the value of the used quantile $Q$.  

Unfortunately, the smallest audience size we obtained from the LinkedIn Ads Manager for $AS(Q,N)$ is 300 due to the privacy-preserving limitation imposed by LinkedIn. In practice, this means that for some value of $N$ referred to as $N_{asymp}$, $V_{AS}(Q)$ will reach an artificial asymptote equal to 300. For any value of $N \geq N_{asymp}$, $V_{AS}(Q)$ will be 300. This prevents us from computing $N_P$ by just using the audience size information retrieved from the LinkedIn Ads Manager since the referred asymptote hides how $V_{AS}(Q)$ decreases until reaching the cutpoint equal to 1.

To reveal the hidden part of the distribution, we fit $V_{AS}(Q)$ using a decreasing exponential model, which uses a logarithmic scale leading to a line, i.e.,  $log(V\textsubscript{AS}(Q)) \simeq -AN + B$  

For this fitting model, we can compute the cutpoint of the regression lines with an audience size equal to one. It is important to note that, as we are using a logarithmic scale model, the cutpoints appear where $log(V\textsubscript{AS}(Q)) = 0$. Therefore, $N\textsubscript{P}$ (the cutpoint) is defined as follows:  $N_P \geq B/A$. We repeat the data aggregation and model fit in 10,000 bootstrap samples, to calculate the 95\% Confidence Interval (CI) for each value of $N$.

The described methodology allows us to estimate $N_P$ for any value of P based on the 50 vectors created from real audiences derived from the  3352 \color{black} users' profiles in our dataset.

\subsection{Skills selection and scenarios} 
 \label{subsec:scenarios}
The way we select skills from the skill set reported by a user in their profile may impact $N_P$. It is very likely that if we combine $N$ rare (unpopular) skills, we get a much smaller audience size than when we combine $N$ very frequent skills.

In this work, we will compare two different strategies. In the first strategy, we randomly select the skills from the skill set. In contrast, the second strategy sequentially adds skills from the least to the most popular. We refer to the former strategy as \textit{Random Selection} and the latter as \textit{Least Popular Selection}.

Overall, we will apply the described methodology in 4 different scenarios:

\begin{itemize}
    \item \pmb{$Sk\_R$}: In this scenario, we only use the user's skills, selected following the random strategy.
    \item \pmb{$Sk\_LP$}: In this scenario, we only use the user's skills, selected following the least popular strategy. 
    \item \pmb{ $Lo\_R$}: In this scenario, we use the reported location and skills of the user, selected following the random strategy.
    \item \pmb{$Lo\_LP$}: In this scenario, we use the reported location and skills of the user, selected following the least popular strategy.
\end{itemize}
{
\section{Users' Uniqueness on LinkedIn and estimated nanotargeting impact}
}
\label{sec:results}

In this section, we first apply the methodology described in the previous section to compute the number of skills that make a user unique on LinkedIn, i.e., $N_P$. Second, based on the $N_P$ results, we compute the success probability of nanotargeting ad campaigns. Finally, using such success probability we estimate the number of LinkedIn users that may be exposed to nanotargeting ad campaigns. 

\subsection{Uniqueness analysis}

Before presenting the results of $N_P$ for the different scenarios  and datasets \color{black}, let us use Figure \ref{fig:sk_r_line_fit} to illustrate how the proposed methodology is applied. The figure shows, in a logarithmic scale, the $V_{AS}(Q)$ vectors for Q = 50, 75, and 90, along with the result of our fitting model. In particular, the figure displays the application of the methodology for the scenario {$Sk\_R\_Agg$ , i.e., location is not used, skills are selected randomly and we use the aggregated dataset}. We note the $Sk\_R$ is the one leading to higher $N_P$ values among all four scenarios considered in our analysis. The results in the figure exhibit that in this scenario, our model does not use $V_{AS}(Q)$ values when $N>30$ since for those values $V_{AS}(Q)$ already collapses in the 300 audience size limit for large percentiles (i.e., $Q$ values). {We provide the same figure for all remaining scenarios and datasets in Appendix \ref{app:scenarios}}. The values of $N$ where $V_{AS}(Q=90)$ collapses to 300 are approximately  24, 14, and 9 \color{black} for the $Sk\_LP\_Agg$, $Lo\_R\_Agg$, and $Lo\_LP\_Agg$ scenarios, respectively. Therefore, our model uses at least  1154 \color{black} audience size samples per $N$ value in all the considered scenarios  for the aggregated dataset \color{black}.

Table \ref{table:pessimistic} shows $N_P$ for $P$ = 50, 75, and 90 {in each scenario ($Sk_R$, $Sk\_LP$, $Lo_R$, $Lo\_LP$, and dataset  (Ds1, Ds2, Ds3, Ds\_Agg)} along with a 95\% confidence interval and the $R^2$ values of the fitting model for each scenario and dataset. The quality of our fitting model based on the reported $R^2$ values seems to be good enough even in the worst case where it is higher than 0.7. 

\begin{table*}[t]
\tiny
\resizebox{\textwidth}{!}{%
\begin{tabular}{c||c|c|c||c|c|c||c|c|c}
$\mathbf{N_P}$ & P=0.5 & 95\% CI & $\mathbf{R^2}$ & P=0.75 & 95\% CI & $\mathbf{R^2}$ & P=0.9 & 95\% CI & $\mathbf{R^2}$\\
    \hline
    Sk\_R\blu\_Ds1  & \blu 19.2 & \blu (18.73,20.39) & \blu 0.91 & \blu 30.8 & \blu (29.18,31.50) & \blu 0.94 & \blu 42.2 & \blu (39.27,45.48) & \blu 0.95 \\ 
    Sk\_LP\blu\_Ds1  & \blu 10.6 & \blu (9.27,12.01) & \blu 0.76 & \blu 28.3 & \blu (26.68,29.84) & \blu 0.80 & \blu 39.1 & \blu (37.44,40.84) & \blu 0.91 \\ 
    Lo\_R\blu\_Ds1  & \blu 8.1 & \blu (6.74,8.26) & \blu 0.87 & \blu 14.2 & \blu (12.91,14.49) & \blu 0.90 & \blu 23.0 & \blu (21.49,24.57) & \blu 0.92 \\ 
    Lo\_LP\blu\_Ds1  & \blu 3.2 & \blu (3.12,3.18) & \blu 0.78 & \blu 6.1 & \blu (6.03,7.72) & \blu 0.74 & \blu 18.2 & \blu (16.38,20.13) & \blu 0.71 \\ 
    \hline
    \blu Sk\_R\_Ds2  & \blu 18.8 & \blu (17.59,19.26) & \blu 0.97 & \blu 28.9 & \blu (27.39,30.60) & \blu 0.96 & \blu 38.7 & \blu (36.71,41.31) & \blu 0.96 \\ 
    \blu Sk\_LP\_Ds2  & \blu 10.0 & \blu (9.78,11.59) & \blu 0.93 & \blu 24.2 & \blu (22.34,25.96) & \blu 0.94 & \blu 35.0 & \blu (32.98,37.65) & \blu 0.97 \\ 
    \blu Lo\_R\_Ds2  & \blu 7.8 & \blu (6.51,7.93) & \blu 0.97 & \blu 12.2 & \blu (10.97,12.39) & \blu 0.96 & \blu 18.2 & \blu (16.77,19.68) & \blu 0.88 \\ 
    \blu Lo\_LP\_Ds2 & \blu 3.1 & \blu (3.11,3.14) & \blu 0.86 & \blu 4.5 & \blu (4.45,5.83) & \blu 0.87 & \blu 10.4 & \blu (8.89,12.02) & \blu 0.76 \\
    \hline
    \blu Sk\_R\_Ds3  & \blu 17.5 & \blu (16.23,17.95) & \blu 0.93 & \blu 25.0 & \blu (23.57,26.38) & \blu 0.96 & \blu 32.9 & \blu (31.45,34.53) & \blu 0.96 \\ 
    \blu Sk\_LP\_Ds3 & \blu 6.9 & \blu (6.76,8.42) & \blu 0.93 & \blu 19.7 & \blu (17.74,22.62) & \blu 0.92 & \blu 30.4 & \blu (28.60,33.90) & \blu 0.98 \\ 
    \blu Lo\_R\_Ds3  & \blu 6.5 & \blu (6.38,7.83) & \blu 0.91 & \blu 10.8 & \blu (10.61,12.23) & \blu 0.90 & \blu 18.0 & \blu (16.56,19.71) & \blu 0.91 \\ 
    \blu Lo\_LP\_Ds3 & \blu 3.1 & \blu (3.08,3.13) & \blu 0.81 & \blu 4.4 & \blu (4.33,4.42) & \blu 0.82 & \blu 8.9 & \blu (7.32,9.33) & \blu 0.77 \\
    \hline
    \blu Sk\_R\_Agg  & \blu 18.6 & \blu (17.56,18.83) & \blu 0.91 & \blu 28.6 & \blu (27.56,29.09) & \blu 0.93 & \blu 38.3 & \blu (36.86,39.63) & \blu 0.95 \\ 
    \blu Sk\_LP\_Agg & \blu 9.1 & \blu (8.95,9.21) & \blu 0.79 & \blu 24.1 & \blu (22.71,24.58) & \blu 0.80 & \blu 35.7 & \blu (34.32,36.94) & \blu 0.89 \\ 
    \blu Lo\_R\_Agg  & \blu 7.9 & \blu (6.60,8.01) & \blu 0.86 & \blu 12.5 & \blu (12.39,12.71) & \blu 0.89 & \blu 21.0 & \blu (19.74,21.45) & \blu 0.89 \\ 
    \blu Lo\_LP\_Agg & \blu 3.1 & \blu (3.12,3.15) & \blu 0.82 & \blu 4.5 & \blu (4.47,5.92) & \blu 0.84 & \blu 14.0 & \blu (12.37,14.26) & \blu 0.71 \\
    \hline
\end{tabular}%
}
\caption{Number of skills needed to identify a user uniquely on LinkedIn with probability 50\% ($N_{50}$), 75\% ($N_{75}$), and 90\% ($N_{90}$), in the four considered scenarios and for the three separated datasets and the aggregated: only using skills selected at random ($Sk\_R\_Ds$), only using skills selected by the least popular selection approach ($Sk\_LP\_Ds$), combining location and skills selected at random ($Lo\_R\_Ds$), and combining location and skills selected using the least popular selection approach ($Lo\_LP\_Ds$). The table also shows the 95\% Confidence Interval (CI) and the R-squared ($R^2$) associated with the fitting model used to obtain $N\textsubscript{P}$}
\label{table:pessimistic}

\Description{Table that shows the results of our analysis. It shows the number of skills needed to identify a user uniquely on LinkedIn with probability 50\% 75\%, and 90\%, in the four considered scenarios and for the three separated datasets and the aggregated: only using skills selected at random (SK_R\_Dsx), only using skills selected by the least popular selection approach (Sk_LP\_Dsx), combining location and skills selected at random (Lo_R\_Dsx), and combining location and skills selected using the least popular selection approach (Lo_LP\_Dsx). The table also shows the 95\% Confidence Interval (CI) and the R-squared associated with the fitting model.}
\end{table*}

{
Let us first discuss the similarity among the results derived from Ds1, Ds2, and Ds3 that use different sampling methods to select profiles. We are especially interested in finding similar results on the scenarios using location (i.e., $Lo\_R$ and $Lo\_LP$) because $>99\%$ of the users in our dataset publicly release their location. Therefore, an attacker will use the location (if it is available) to run nanotargeting ad campaigns.

For P=50\%, the three datasets show comparable results in most of the scenarios. For the scenario $Lo_R$, $N_{50}$ is equal to 8.1, 7.9 and 6.5  for Ds1, Ds2 and Ds3, respectively. For the scenario $Lo_{LP}$ we observe almost identical $N_{50}$ results ranging between 3.1 and 3.2 in the three datasets. If we now observe the results for P=75\%, we obtain a similar conclusion. The results are comparable across the three datasets. For $Lo_R$, $N_{75}$ ranges between 10.8 (Ds3) and 14.2 (Ds1), with only 3 skills of difference. The difference across the three datasets reduces to only 1.5 skills if we consider the $Lo_{LP}$. Finally, for the $P=90$, we observe a rather reasonable difference for the $Lo\_R$ scenario where the estimated $N_{90}$ value for Ds1, Ds2, and Ds3 is 23, 18.2, and 18. However, for the $Lo\_{LP}$ we observe a rather large difference of 9 skills between Ds1 ($N_{90}=18.2$) and Ds3 $N_{90}=8.9$. Even if we accept as valid the most conservative result, i.e.,  $N_{90}=18.2$, the portion of LinkedIn users with more than 18 skills is around 45\%. This means that almost 1/2 of the LinkedIn users may be affected by this lower bound case in the $Lo\_LP$ scenario. Overall, when we consider the location, we find comparable results among the three datasets except Ds1 for $N_90$ which is twice the value obtained from Ds2 and Ds3. 

Results' consistency across the three datasets proves robustness, affirming the feasibility of uniquely identifying LinkedIn users through public non-PII from their profiles (location and professional skills). From now on in the paper, unless otherwise stated, we will use the results of the aggregated datasets for our analysis and discussion.

To conclude this subsection, we discuss the main results we can extract from the analysis of LinkedIn users' uniqueness based on non-PII out of applying the proposed methodology to the aggregated dataset. 

First, as we expected, the results in the table show that using the location significantly reduces the number of skills required to identify users on LinkedIn uniquely. This reduction is roughly 2$\times$ for the random skill selection (i.e., $Sk\_R$ Vs. $Lo\_R$) and 3$\times$-4$\times$ for the least popular selection (i.e., $Sk\_LP$ Vs. $Lo\_LP$).  For the two scenarios that do not consider the user location, at least 24 skills are required to make a user unique with a probability $\geq75\%$. If we look again at Figure \ref{fig:CDF_skills_per_user}, we observe that around 30\% of the users report 25 or more skills in their profile. This means that exclusively using the skills roughly reduces the nano-targetable users on LinkedIn to 1/3 of its user base. It is worth noting that this still represents a privacy risk for {$\sim$323M} LinkedIn users. However, in practice, the advertiser (attacker) willing to target an individual will be able to use the location in the vast majority of the cases ($\sim$99\%), which reduces a lot the number of skills required to successfully nanotarget the user.

Second, based on the $Lo\_R\_Agg$ scenario results, 8, 13, and 21 random skills are enough to make a user unique on LinkedIn with a probability of 50\%, 75\%, and 90\%, respectively. When we focus on the $Lo\_LP\_Agg$, the number of rare skills that make a user unique with those same probabilities is 3, 5, and 14\, respectively. 

In summary, according to the aggregated dataset, the combination of the location and 5 rare skills reported is enough to uniquely identify 3 out of 4 LinkedIn users. Increasing the number of skills to roughly 21 would allow uniquely identifying 9 out of 10 users.
}

\subsection{Success probability of nanotargeting campaigns}

{ $N_P$ is a metric that measures LinkedIn's user uniqueness but it is also valid to estimate the success probability of a nanotargeting campaign. The success of a nanotargeting campaign implies being able to create an audience size equal to 1 for the targeted user (i.e., making that user unique). Once an attacker can define such an audience, if the platform allows configuring and running the campaign, they may be able to deliver nanotargeted ads to the targeted individual. Therefore, $N_P$ allows establishing the probability of success of a nanotargeting campaign. Given the number of skills $N_{attacker}$ known by the attacker, we just need to find the value of $P$ where $N_{attacker}~=~N_P$. That $P$ will be the success probability of the attack. Therefore, $N_P$ allows us to define the probability of success of a nanotargeting ad campaign.}

\begin{figure}[t]
    \includegraphics[width=\hsize]{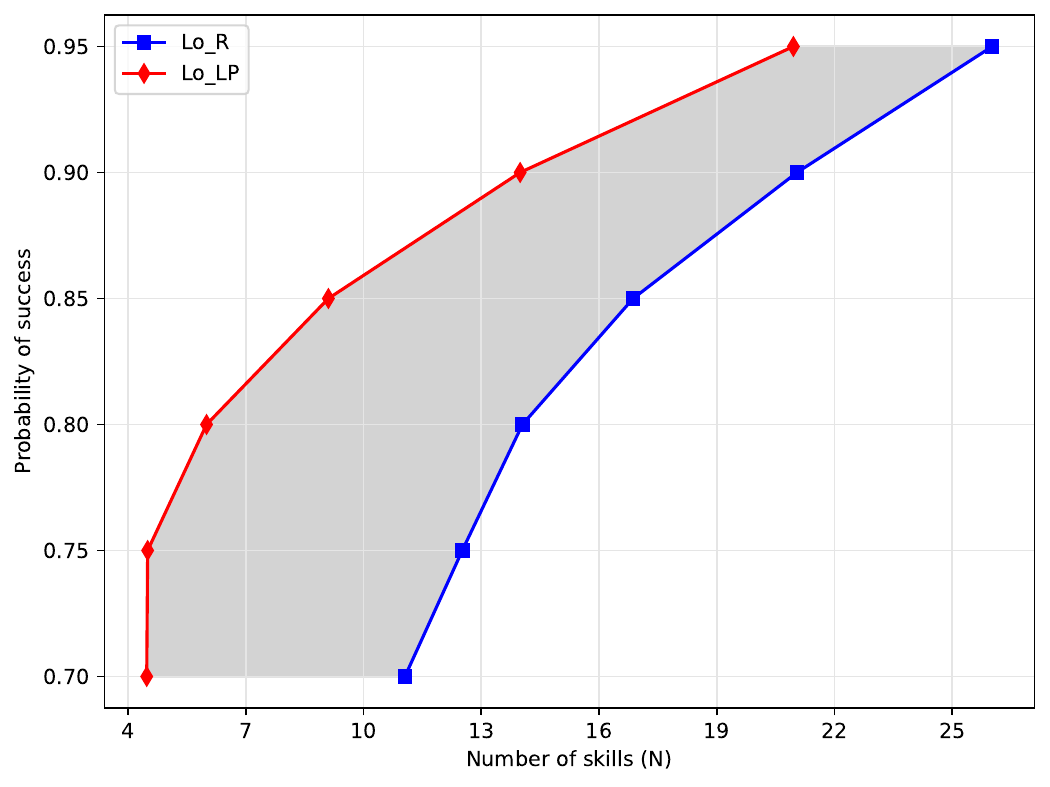}\hfill
	\caption{Probability of success of a nanotargeting campaign by combining the location and $N$ skills. The red line represents an upper bound linked to using the least popular selection strategy for skills ($Lo\_LP\_Agg$). The blue line represents a lower bound linked to using the random selection strategy for skills ($Lo\_R\_Agg$).}
	\label{fig:real_attack_success_probability}
    \Description{Line graph showing the estimated uniqueness probability/nanotargeting success ratio for a user with N skills listed. The X axis represents the number of skills listed by the user, while the Y axis represents the probability of uniqueness. There are two lines, each one corresponds to the estimations made by each of the following scenarios: using location and skills selected at random (Lo_R_Agg) and using location and skills selected by the least popular (Lo_LP_Agg). the Lo_LP scenario estimates around 4 to 6 skills less to reach the same uniqueness probability.}
\end{figure}

Figure \ref{fig:real_attack_success_probability} shows the expected success probability (y-axis) of a nanotargeting campaign based on the combination of the location and $N$ skills (x-axis) { and the results of our model applied to the aggregated dataset}. The figure depicts an upper bound for the success probability (red line) computed for the case where the least popular skills are selected and a lower bound (blue line) that refers to the random selection approach. The results suggest that an advertiser would need to use (roughly) between 5-8 skills more to achieve the same success probability when applying the random skills selection instead of the least popular skills selection. %This is true except for very high success probabilities, such as 95\%, where both strategies require a very similar number of skills. 

The success probability reported in this section is a reference extracted from our dataset. From a practical point of view, an advertiser willing to nanotarget an individual would use all the skills the individual reports in her profile because the cost of retrieving 1 or 40 skills is the same. If an advertiser selects all the skills for an individual, the random and least popular skill selection becomes the same strategy. 

We want to highlight that Figure \ref{fig:real_attack_success_probability} can estimate the success probability of a nanotargeting campaign that uses all the skills available in the user profile. We just need to find the success probability for the number of skills $N$ used in the campaign. For this case, we recommend using the success probability of the random approach which is more realistic (mix of skills of different audience sizes). Therefore, any user can estimate the risk of being successfully targeted through a nanotargeting ad campaign based on the skills they publicly report in their profile. Therefore, the success probability of a nanotargeting campaign is bounded by the amount of skills users report in their profiles. The more skills users report the more vulnerable they are to nanotargeting attacks.

Finally, it is worth noting that the success of a nanotargeting attack on a specific individual also depends on the audience size of the location and skills reported. For instance, it would be easier to nanotarget a user reporting as location Brussels Metropolitan Area (audience size 1.4M) than a user reporting as location United States (audience size 220M). Similarly, users reporting weird skills would be easier to nanotarget than users that only report very popular skills. Section \ref{sec:dataset} shows that some skills are reported by less than 1k users on LinkedIn while other skills are reported by more than 100M users. The former skills make users more nanotargetable. In summary, how easy nanotargeting a specific user on LinkedIn depends on the number of skills they report, but also on how singular they are. However, the bug fix implemented by LinkedIn eliminates the risk of nanotargeting in their platform, so these results currently apply to assessing the uniqueness of users on LinkedIn, and ads can still be delivered to the targeted users, but the ad may not exclusively reach the targeted user as the targeted audience must include other 299 users.

\color{black}

{

\subsection{Estimation of the number of LinkedIn users potentially exposed to nanotargeting ad campaings}

We complete this section by estimating the number of LinkedIn users that may have been exposed to nanotarteting ad campaigns. We adopt a practical approach and consider attackers would use all available skills in the targeted user profile.  We could use two different success probabilities for the number of skills employed according to Figure \ref{fig:real_attack_success_probability}, the one associated with using the least popular skills (upper bound) and the one using skills at random (lower bound). We decided to use the success probability associated with the random procedure that is more realistic in a potential real attack in which the attacker would very likely use all the skills (mix of skills with low, medium, and large audience size) the targeted user has reported in their profile.

We estimate the number of impacted users for different values of success probability $P$. For a given $P$ value, we first compute $N_P$ and the portion of LinkedIn users that has reported $\leq N_P$ skills in their profile (according to the aggregated dataset). We then multiply $P$ by that portion of users and by the total number of users on LinkedIn (i.e., 970M).  

Table \ref{table:total_users_affected} shows the estimated number of LinkedIn users susceptible to being nanotargeted for different $P$ values ranging between 0.7 and 0.95. From a privacy perspective, the worst case is the one where more users are exposed to the potential nanotargeting attack. That scenario happens for the value $P=0.8$ (14 skills required) where the number of users that may be potentially nanotargeted is 419M, which represents 43\% of the LinkedIn user base. The best possible scenario among the ones depicted in the table is $P=0.95$ (26 skills required). Still, in this case, the number of users at risk was 258M, 26.5\% of the LinkedIn user base. 

In a nutshell, independently of which value we retrieve from the table, our estimations reveal that, even in the most optimistic scenario, a quarter billion users were nanotargetable on LinkedIn with hyper-tailored ad creativities before LinkedIn fixed this issue.

}

\begin{table}[t]
\centering
\small
\begin{tabular}{c c c c c c c}
        \toprule
        \blu Probability & \blu Nº of skills & \blu Portion & \blu Total & \blu LinkedIn\\
        \blu of &  \blu (N) needed to & \blu of LinkedIn &  \blu number of  & \blu  users  \\
        \blu user & \blu achieve that  & \blu users listing & \blu users & \blu affected\\
        \blu uniqueness & \blu probability & \blu $\geq$N skills & \blu affected (M) & \blu (\%)\\
        \midrule
        \blu 0.7 &  \blu 11 & \blu 0.59 & \blu 400,61 & \blu 41.3\%\\
        \midrule
        \blu 0.75 &  \blu 13 & \blu 0.56 & \blu 407,4 & \blu 42.0\%\\
        \midrule
        \blu 0.8 &  \blu 14 & \blu 0.54 & \blu 419,04 & \blu 43.2\%\\
        \midrule
        \blu 0.85 &  \blu 17 & \blu 0.47 & \blu 387,52 & \blu 40.0\%\\
        \midrule
        \blu 0.9 & \blu 21 & \blu 0.37 & \blu 323,01 & \blu 33.3\%\\
        \midrule
        \blu 0.95 & \blu 26 & \blu 0.28 & \blu 258,02 & \blu 26.5\%\\
        \bottomrule
\end{tabular}
\caption{Estimations of unique users according to our analysis results of the aggregated dataset. The first column shows the probability of user uniqueness, the second column shows the range of the number of skills (N) needed to be unique with that probability (taken from the estimations of the Lo\_R scenario of the aggregated dataset), the third column shows the portion of users in our aggregated dataset that share N skills or more in their profiles. The estimations, in the two last columns, are calculated by multiplying the total number of users on LinkedIn by the probability of being unique with N skills (taken from Figure \ref{fig:real_attack_success_probability}) and by the proportion of users that report N skills or more in our aggregated dataset. We estimate that the number of users that are unique with their location and skills on LinkedIn is between 258M and 419M, in other words, between 26.5\% and 43\% of the LinkedIn users.}
\label{table:total_users_affected}
\Description{Estimations of unique users according to our analysis results of the aggregated dataset. The first column shows the probability of user uniqueness, the second column shows the range of the number of skills (N) needed to be unique with that probability (taken from the estimations of the Lo\_R scenario of the aggregated dataset), the third column shows the portion of users in our aggregated dataset that share N skills or more in their profiles. The estimation, in the last column, is calculated by multiplying the total number of users on LinkedIn by the probability of being unique with N skills (taken from Figure \ref{fig:real_attack_success_probability}) and by the proportion of users that report N skills or more in our aggregated dataset. We estimate that the number of users that are unique with their location and skills on LinkedIn is between 258M and 419M, in other words, between 26.5\% and 43\% of the LinkedIn users.}
\end{table}
\section{Nanotargeting proof of concept}
\label{sec:poc}

If our model outcome is correct, it may be possible to nanotarget an individual on LinkedIn. By nanotargeting, we refer to showing ads from an ad campaign exclusively to the targeted individual. LinkedIn claimed, at the time we implemented our proof of concept experiment, that it was not possible to launch ad campaigns for audience sizes $<$300 users. If LinkedIn had effectively imposed this policy, we would not have been able to run nanotargeting campaigns. In a nutshell, in this section, we verify whether it was feasible to run nanotargeting campaigns on LinkedIn based on the results derived from our methodology.

\begin{figure}[t]
    \includegraphics[width=\hsize]{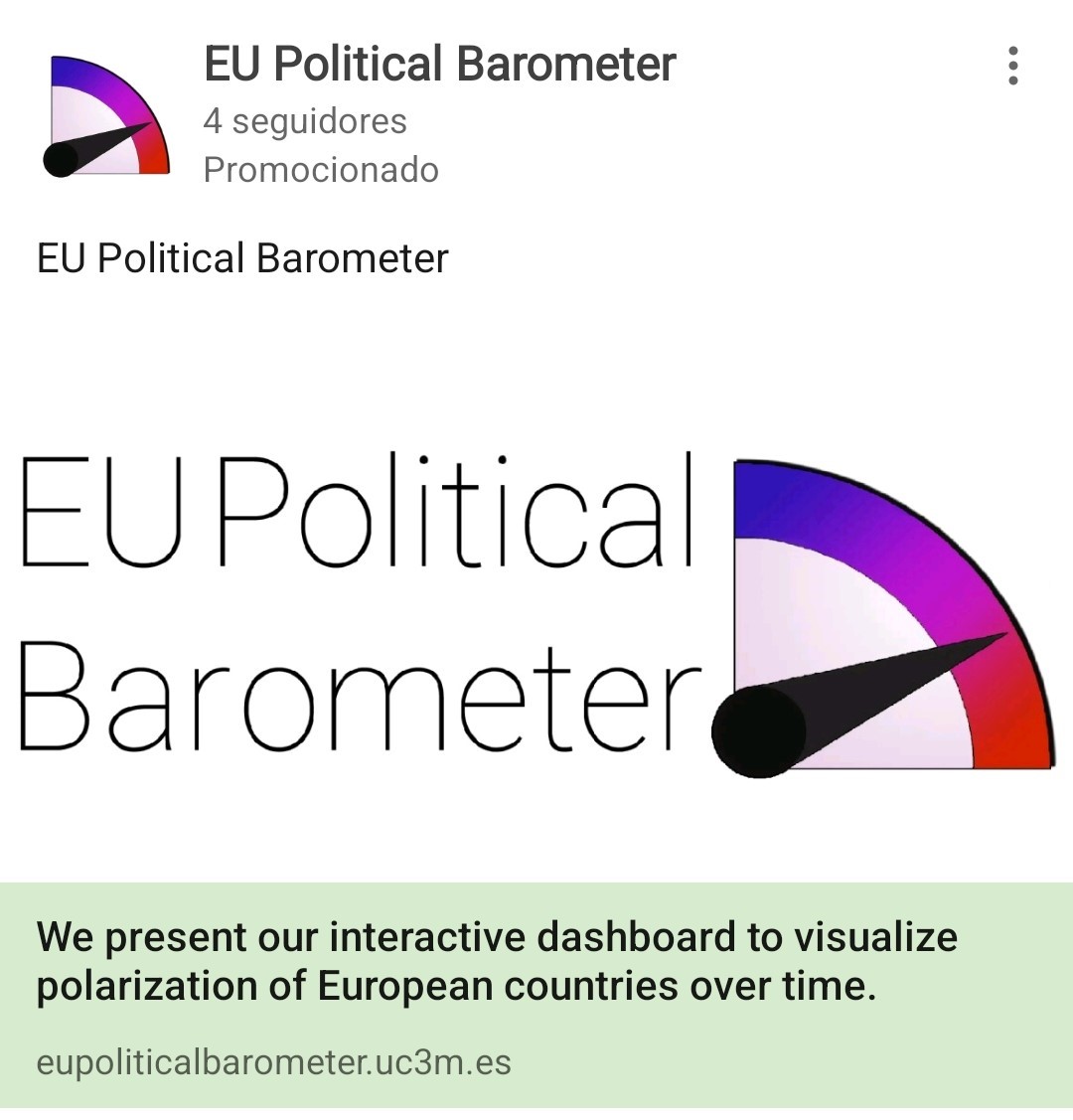}
    \caption{Ad creativity used in the proof of concept experiment.}
    \label{fig:pocad}
    \Description{Image showing the ad creativity we used in our proof of concept experiment.}
\end{figure}

\subsection{Description of the experiment}
We aimed to nanotarget three of the authors of this paper based on their self-reported location and skills. From now on, we will refer to the authors as user 1 (U1), user 2 (U2), and user 3 (U3), respectively. 

To configure each campaign, we used the LinkedIn Campaign Manager and defined the targeted audience using the location and N skills retrieved from the LinkedIn profile of the targeted user. In addition, we set up the budget, uploaded the ad creativity, and defined the landing page the users would visit if they clicked on our ads.

Once a LinkedIn advertising campaign is defined, to continue with its publication and be launched to the public, LinkedIn offers the possibility to use two different buttons including the text \textit{"Launch Campaign"}, one at the right side of the page, and another at the bottom, that is only visible if the advertiser scrolls down. We could select either of these buttons to publish the ad. 

In our nanotargeting campaigns, we observed that the "Launch Campaign" button on the right side of the Ads Manager was not clickable, arguing that the audience was too small. However, this measure could be bypassed using a simple JavaScript code in the browser's console to reactivate the button: 

$document.querySelector(button\_selector).disabled = false.$

At first, we thought LinkedIn was implementing its policy to avoid ad campaigns targeting less than 300 users. However, after enabling the button, the campaign could be launched and the audience size was not checked during the ad review process. {  In November 2023 we discovered LinkedIn had fixed this bug and was properly enforcing its policy of not allowing campaigns with an audience size below 300 users. This is a very positive step that is further discussed in Section \ref{subsec:fix}}

Next, we detail each of the campaign attributes that were relevant to our proof of concept experiment. 

\noindent\textit{\textbf{Skills selection}:} The number of skills available in the profiles of the targeted individuals was 28, 42, and 28 for U1, U2, and  U3, respectively. Our model's results enabled us to choose any of the two potential skill selection strategies: random or least popular. We decided to run our proof of concept experiment by selecting skills at random. This is to emulate the simplest setting for a non-skilled advertiser willing to implement a nanotargeting campaign. As we explained, any user (advertiser) with a LinkedIn account can retrieve the skills reported by any other user.  It is enough to access the profile, retrieve the skills and the location reported by the user being targeted, and configure an ad campaign in the dashboard using that information. In contrast, implementing the least popular selection implies sorting the skills by popularity, which requires accessing the Ads Manager and obtaining the audience size associated with each skill. Although this is a very simple step for savvy users, non-skilled users may not know how to obtain the audience size for each skill and will be unable to implement the least-popular skill selection in the nanotargeting campaign.

\noindent\textit{\textbf{Number of skills}:} We configured campaigns with 7, 10, 13, 16, and 19 randomly selected skills along with the location. { We did not run campaigns using all the skills available in the profile (as an attacker would do) because we aim to validate if our estimated nanotargeting success probabilities are accurate. Success in any nanotargeting ad campaign implies that one with more skills would have succeeded as well.}

\noindent\textit{\textbf{Campaign duration}:} All campaigns ran for 3 days (72 hours). Each campaign started at day d noon and finished at d+3 noon. We note that the starting day, d, was not the same for all the campaigns. 

\noindent\textit{\textbf{Campaign budget}:} Each campaign was configured with a budget of \$10. None of the 15 ad campaigns spent the budget in the 3 days they were running.

\noindent\textit{\textbf{Ad creativity}:} We used a neutral ad creativity advertising a website from a research project that has nothing to do with privacy. Figure \ref{fig:pocad} shows the ad creativity employed in all our ad campaigns. 

\noindent\textit{\textbf{Targeted device}:} We configured our campaigns to deliver ads both on mobile devices and desktops. 

Overall,  we were targeting 3 different users and we were running 5 campaigns for each of them (one per number of skills value). Therefore, our proof of concept experiment included 15 nanotargeting campaigns in total. Table \ref{table:expected_probs_poc} shows for each value of skills (first column) the estimated success probability according to our model (second column) and the expected number of successfully nanotargeted users among the 3 targeted users (third column). We compute the latter by multiplying the success probability retrieved from our model by the number of campaigns run per skills value, i.e., 3. For instance, for 19 skills (89\% success rate) the expected number of successful campaigns out of three launched campaigns, based on the results of our methodology, was 2.67. This implies that at least 2 and likely 3 out of the three campaigns using 19 skills should be successful in our experiment. The last column of the table shows the actual number of successful nanotargeting campaigns in our experiment.

\begin{table}[t]
\centering
\begin{tabular}{c c c c c c c}
        \toprule
         & & Model & Experiment \\
        Skills & Model probability & nanotargeted &  nanotargeted \\
        & & campaigns & campaigns\\
        \midrule
        7 &  \blu 0.49 &  \blu 1.47 & 1\\
        \midrule
        10 &  \blu 0.66 & \blu 1.98 & 2\\
        \midrule
        13 &  \blu 0.77 & \blu 2.31& 3\\
        \midrule
        16 &  \blu 0.84 & \blu 2.52 & 3\\
        \midrule
        19 &  \blu 0.89 & \blu 2.67 & 3\\
        \bottomrule
\end{tabular}
\caption{Expected and actual successful nanotargeting campaigns in the proof of concept experiment. The first column includes the skills used in the campaign. The second column shows the success probability retrieved from the applied methodology. The third column shows the expected number of successful campaigns in the experiment out of the three targeted users per number of skills. The fourth column shows the actual number of successful campaigns in the proof of concept experiment.}
\label{table:expected_probs_poc}
\Description{Table that shows the expected and actual successful nanotargeting campaigns in the proof of concept experiment. The first column includes the skills used in the campaign. The second column shows the success probability retrieved from the applied methodology. The third column shows the expected number of successful campaigns in the experiment out of the three targeted users per number of skills. The fourth column shows the actual number of successful campaigns in the proof of concept experiment.}
\end{table}

\subsection{Validation of nanotargeting success}

To validate whether our campaigns had successfully nanotargeted the targeted individual, we relied on both the information provided by LinkedIn for our campaigns and the information we directly collected. 

First, we used the information provided by LinkedIn to advertisers in a dashboard where they can monitor the progress of their campaigns. It delivers information for many parameters, including the number of impressions and the number of clicks for an ad campaign. In some cases, it also estimates the (unique) users reached (referred to as reach estimation) in the campaign. This last parameter would allow us to confirm the success of the nanotargeting campaign when it is equal to 1 once the campaign is over. However, this parameter presents two limitations: $(i)$ LinkedIn informs that this parameter is in a beta version and it only offers an estimation; $(ii)$ we observed that the estimation is only available in those campaigns reaching multiple users, but it is never reported when very few users are reached. Therefore, while we report this value (see Figure \ref{fig:ldrep} in Appendix \ref{app:LinedIn_records}), we could not rely on it to verify the success of a nanotargeting campaign, but the opposite when the campaign had reached multiple users.

Second, all the targeted authors were aware of the ad creativity we were using in the ad campaigns, and we instructed them to $(i)$ take a snapshot of each ad impression received from the nanotargeting campaign; $(ii)$ click on the nanotargeted ad every time it appeared in their LinkedIn feed.\footnote{We note U3 forgot to click in one of the received ad impressions in the campaign using 13 skills (marked with * in table \ref{table:poc_results}). In that case, as we will find in our results and the LinkedIn report, the campaign delivered 3 ad impressions and received 2 clicks.} When clicking on the ad, the user was forwarded to the research project website advertised that runs on a server we manage. The server recorded the timestamp for each click and the campaign from which the click was generated, which identifies the user (U1, U2, or U3) performing the click. { We note we neither collect the IP address of the user nor any other personal ID beyond the campaign ID to certify which user had received the ad.}

With the information obtained in the two previous steps, we could assess whether a nanotargeting campaign was successful. We could confidently conclude that the user was the only one who received the ad if the number of impressions and clicks reported by LinkedIn matched the number of impressions and clicks provided by the targeted users and the number of clicks logged in our backend system, where we can verify whether the clicks comes from a single user.

\subsection{Results of nanotargeting experiment}

Table \ref{table:poc_results} shows the results from the 15 ad campaigns we ran in our proof of concept experiment. For each campaign, the table identifies: $(i)$ the user being targeted, $(ii)$ the number of skills used in the campaign, $(iii)$ the number of impressions and clicks reported by LinkedIn in the dashboard summarizing the campaign results, $(iv)$ the number of impressions reported by the user through the snapshot they captured of the received ads, $(v)$ the number of clicks registered in our backend server, and $(vi)$ the cost of the campaign. We highlight in bold all campaigns that successfully nanotargeted the targeted individual. Figure \ref{fig:ldrep} in Appendix \ref{app:LinedIn_records} shows a snapshot of the results of our campaigns as reported in the LinkedIn dashboard.   

{ We note that the targeted users are not statistically easier to nanotarget than a median user from our dataset. The audience size associated with the (same) location of the three targeted users was 4.2M, while the median audience size for the locations in our dataset is 1M. Similarly, the median audience size for the skills reported by U1, U2, and U3 is 10.5M, 21M, and 11.75M, respectively. This is 10-20$\times$ higher than the median audience size for the skills in our dataset, i.e., 1.2M.}

All the campaigns using the 13, 16, and 19 skills (along with the location) successfully nanotargeted the targeted user. Also, 2 out of the 3 campaigns using 10 skills were successful. Finally, only one of the campaigns using 7 skills was successful. These results match the expectations derived from our model as reported in Table \ref{table:expected_probs_poc}. Our intuition is that our model provides a conservative result, and the actual success probability would be slightly higher than what our model reports. This intuition is based on the fact that using 13 skills (77\% success probability) already led to successful nanotargeting campaigns in all the cases. 

The primary outcome of this experiment is that we have demonstrated that running nanotargeting campaigns systematically on LinkedIn was feasible at least until July 2023. To the best of our knowledge, this experiment is the first demonstration that the combination of publicly available non-PII data can be used to individually target users. From the research point of view, the bug fix implies that our study cannot be reproduced.

\begin{table}[t]
\begin{minipage}{\columnwidth}
\begin{center}
\small
\begin{tabular}{l c c c c c c c c c c c}
\toprule
\multicolumn{1}{c}{\multirow{2}{*}{UID}} & \multicolumn{1}{c}{\multirow{2}{*}{Skills}} & \multicolumn{2}{c}{LinkedIn Report} & User Report & Backend &  \multicolumn{1}{c}{\multirow{2}{*}{Cost (\$)}}\\
\cmidrule{3-6}
& & Imps & Clicks & Impressions & Clicks & \\

\midrule
\multicolumn{1}{c}{\multirow{2}{*}[-20pt]{1}}   &  7        & 63        & 4         & 3         & 4         & 6.57      \\

                                                & 10        & 55        & 2         & 2         & 2         & 4.71      \\
                                          
                                                & \bk{13}   & \bk{3}    & \bk{3}    & \bk{3}    & \bk{3}    & \bk{0.20} \\

                                                & \bk{16}   & \bk{3}    & \bk{3}    & \bk{3}    & \bk{3}    & \bk{0.20} \\

                                                & \bk{19}   & \bk{3}    & \bk{3}    & \bk{3}    & \bk{3}    & \bk{0.10} \\
\midrule

\multicolumn{1}{c}{\multirow{2}{*}[-20pt]{2}}   & \bk{7}    & \bk{2}    & \bk{2}    & \bk{2}    & \bk{2}    & \bk{0.11} \\

                                                & \bk{10}   & \bk{1}    & \bk{1}    & \bk{1}    & \bk{1}    & \bk{0.16} \\

                                                & \bk{13}   & \bk{2}    & \bk{2}    & \bk{2}    & \bk{2}    & \bk{0.15} \\

                                                & \bk{16}   & \bk{3}    & \bk{3}    & \bk{3}    & \bk{3}    & \bk{0.14} \\

                                                & \bk{19}   & \bk{3}    & \bk{3}    & \bk{3}    & \bk{3}    & \bk{0.65} \\
\midrule
                                                    
\multicolumn{1}{c}{\multirow{2}{*}[-20pt]{3}}   & 7         & 85      & 2           & 2         & 2         & 8.35      \\

                                                & \bk{10}   & \bk{3}    & \bk{3}    & \bk{3}    & \bk{3}    & \bk{0.39} \\
                                                
                                                & \bk{13}   & \bk{3}    &\bk{2*}    & \bk{3}    & \bk{2*}   & \bk{0.69} \\

                                                & \bk{16}   & \bk{3}    & \bk{3}    & \bk{3}    & \bk{3}    & \bk{0.52} \\

                                                & \bk{19}   & \bk{3}    & \bk{3}    & \bk{3}    & \bk{3}    & \bk{0.56} \\
                             
\bottomrule                                            
\end{tabular}
\end{center}
\caption{Results of the proof of concept experiment. Under LinkedIn Report, the results reported by the LinkedIn Campaign Manager; under user report, the impressions the user notified for each campaign; and under Backend Log, the ad clicks recorded in our backend server for each campaign. It also includes the cost of each campaign in USD.}
\label{table:poc_results}
\Description{Table that shows the results of the proof of concept experiment. For each participant, it shows the different campaigns launched, one per row, with a different number of skills used to define the audience. For each campaign, the table shows the impressions and clicks reported by LinkedIn for that campaign, the number of impressions reported by the targeted user, the number of clicks registered in our backend, and the cost of that campaign.}
\end{minipage}
\end{table}  
\section{Discussion}
\label{sec:discussion}

In this paper, we have shown that nanotargeting could be implemented systematically on LinkedIn before the bug got fixed. The only requirement was having an active LinkedIn account, learning how to run an ad campaign on LinkedIn, and activating the disabled button to launch the campaign. In this section, we discuss some issues derived from our work.

\subsection{Theoretical privacy limits and awareness}

Our work contributes to the existing body of literature that has demonstrated in a different context that our privacy is bounded by a handful of non-PII attributes. In the case of LinkedIn, the location and {\blu 5} rare skills make a user unique with a probability of 75\%. Our results reinforce the undeniable fact that our privacy is a very vulnerable asset.

Unfortunately, our intuition is that users have an unconscious safety feeling when they share non-PII data. Talking to some computer science colleagues (not necessarily working in privacy) about our research, most of them found added value to reporting professional skills on their LinkedIn profile since that would allow other colleagues to know their expertise. However, none considered sharing that information may represent a risk. They were very surprised when we explained that they could be nanotargeted with ads based on the professional skills they publicly report.

Although it is not a scientific experiment, we extract two important lessons from those informal discussions with our colleagues. First, if skilled users struggle to identify potential privacy risks of sharing non-PII data, that means there is a lot of work ahead for making regular Internet users aware of how vulnerable they are, even if they are careful not to share and protect their PII data. Second, non-PII data may expose users to a challenging dichotomy. On the one hand, users may find value in sharing non-PII data, such as in the case of professional skills. On the other hand, the more non-PII data they share, the more vulnerable they become in terms of privacy, as our results have demonstrated.

\subsection{Nanotargeting on LinkedIn}

First, we probed that it was straightforward to configure a nanotargeting campaign on LinkedIn since all the required information is available in the targeted user's profile. In practice, an attacker would likely use all the skills in the targeted user's profile since gathering one or many skills is equally simple. This means the nanotarteging risk a user on LinkedIn was exposed to was directly proportional to the number of professional skills they reported in their profile. 

Second, the cost of nanotargeting campaigns was very low. The cost of the successful nanotargeting campaigns in our experiment ranged between \$0.10 and \$0.69. This means that implementing a nanotargeting campaign was roughly free. 

Third, although we ran the campaigns for only three days, in all but one campaign, we obtained at least 2 impressions. On average, users were exposed to 2.67 ad impressions in successful nanotargeting campaigns. Roughly speaking, we impacted the targeted users once a day. More importantly, the experience of the three targeted authors was that when the nanotargeted ad was displayed, it usually was the first ad impression in the user's newsfeed. Therefore, they did not have to scroll down to visualize the nanotargeted ad impression. All these elements suggest that it is feasible to frequently expose the targeted individual to tailored content embedded in the nanotargeted ads in case they are active on LinkedIn. 

In a nutshell, our proof of concept experiment suggested that it was feasible to continuously expose users to nanotargeted ads at an extremely low cost on LinkedIn, including very influential users, before LinkedIn fixed this issue.

\subsection{Nanotargeting influential people}

LinkedIn was especially interesting since it allows for easy access to profiles of relevant and influential people worldwide, such as politicians, CEOs, etc. Therefore, our research opened a question regarding how easy it would have been to send hyper-personalized ads to these users. We gathered the location and number of skills from 120 LinkedIn Top Voices 2020 \cite{ldtopvoices} as a sample of LinkedIn influential users to check how many of them could be exposed to nanotargeting according to our results. All of them report a location except one (Richard Branson). The number of skills in their profiles is 28 on average, and 94 of them publish 13 skills or more. Based on our results, most of these users were susceptible to being reached with LinkedIn nanotargeting campaigns. 

\subsection{Legal discussion}
Modern data protection regulations such as the General Data Protection Regulation (GDPR) \cite{gdpr}, enacted in all EU countries in May 2018, eliminate the PII concept to avoid the misconception that personal data refers only to PII items. Instead, Article 4 of the GDPR includes the following definition: \textit{‘personal data’ means any information relating to an identified or identifiable natural person (‘data subject’)}. Although we are not legal experts, it is clear from the definition that the combination of skills and location on LinkedIn falls under the definition of personal data.
 
While the text of the GDPR is clear regarding what personal data is, in practice, it is tough to validate when a specific combination of non-PII items makes a user unique in an online application. For instance, in most cases, there is no way an honest advertiser on LinkedIn (or other applications) can assess whether a particular audience created through the combination of non-PII attributes is personal data. 

In our opinion, it is urgent to concretize the concept of personal data in the context of non-PII. To this end, the research community should work with data protection authorities to create guidelines informing companies when they should treat combinations of non-PII attributes as personal data.

\subsection{LinkedIn's reaction to responsible disclosure}

We considered that our research was unveiling an obvious privacy vulnerability within the LinkedIn Advertising Platform. Even more, there were security implications as nanotargerting could also compromise the users' device (e.g., through malvertising attacks). Furthermore, our work also proved that LinkedIn failed to implement its policy that limited the targeted audiences of ad campaigns to $\geq$300 users \cite{ldhelpaudience}.  Therefore, we believe our research unveiled a privacy and security vulnerability within the LinkedIn Ads platform.

Based on the standard practice in the security community, we followed a responsible disclosure based on the guidelines provided by LinkedIn. We contacted LinkedIn to make them aware of the unveiled vulnerability and give them the opportunity to fix it before our work becomes public. 

LinkedIn's Security Page \cite{ldsecurity} refers to their bug bounty program on HackerOne as the channel to report system vulnerabilities. We found no other procedure to report privacy or security issues on the LinkedIn platform. We submitted a report describing our findings, but at the triage stage, managed by HackerOne, they considered that the issue we reported was out of the scope according to LinkedIn's Policy Page within the platform \cite{ldhackerone}, because the bug we reported requires social engineering to implement it. Contrarily, we believe it was the other way around. The bug allowed attackers to run social engineering attacks, for instance, to manipulate users through hyper-personalized ads. The bug existed independently of the potential nanotargeting attacks because running ad campaigns to target audiences involving less than 300 users was possible. Running such a campaign does not require any social engineering.

The vulnerability we disclosed in this work $(i)$ implied privacy and security risks for LinkedIn users, $(ii)$ it could imply that LinkedIn was not complying with current data protection regulations such as the GDPR. We are glad that finally, as explained in subsection \ref{subsec:fix}, LinkedIn has implemented their ad policy correctly, despite they ignored our report in the first place.
%It is extremely surprising to us that LinkedIn ignored in the first place the vulnerability we disclosed in this work since $(i)$ it implied privacy and security risks for LinkedIn users, $(ii)$ it could imply that LinkedIn was not complying with current data protection regulations such as the GDPR. 

{
\subsection{LinkedIn's fix of nanotargeting vulnerability}
\label{subsec:fix}
We informed LinkedIn of the vulnerability in April 2023, and they considered our research was not unveiling a security/privacy vulnerability. We tested that nanotargeting was feasible on LinkedIn at least until July 2023 when we implemented our last nanotargeting campaign. As part of a review process in ACM CHI 2024, in November 2023 we tried to run a new nanotargeting campaign. Then, we verified LinkedIn had fixed the reported bug making it impossible to run campaigns targeting audiences lower than 300 users. Therefore, running nanotargeting campaigns is not possible anymore on LinkedIn, but it does not prevent attackers from sending ads to the targeted user. However, with the current set up you need to define an audience including at least 300 users. That means, an attacker may still reach the targeted user, but the ad will be delivered to some other users as well.

As explained in Section \ref{sec:poc}, formerly, anyone could launch a campaign with an audience size $<$300 by enabling the disabled button and clicking it, without any server-side validation of this policy. The newly created campaign used to go to \textit{in review} status and eventually became active. Now, if anyone follows these steps, the newly created campaign remains in a new status \textit{audience too small}, and there is no way of activating it without modifying the audience size.

%The ultimate goal of our research was to contribute to protecting the privacy of hundreds of millions of users on LinkedIn. In other words, our objective was to get LinkedIn to fix the reported vulnerability. 
We aimed to protect LinkedIn users' privacy, urging LinkedIn to address the vulnerability. Therefore, we are very happy LinkedIn reacted to our research by fixing it. It is unfortunate LinkedIn did not give us any credit for this, but we are equally happy because our goal has been achieved.

%Although someone might think it was unfortunate for us this happened before our research was accepted for publication because it may reduced the relevance of our research, we do not share that point of view. We believe it is extremely important to publish success stories around academic works in the area of privacy to demonstrate it is feasible to achieve practical positive impacts through research. In our opinion, the ultimate goal of any privacy researcher should be having a real impact on improving the privacy/security of the users. 
}

\subsection{Countermeasures to prevent nanotargeting}

The most efficient measure to preclude advertisers from running nanotargeting ad campaigns is implementing a policy, like the one LinkedIn advertises \cite{ldhelpaudience}, that establishes a minimum audience size to allow running a campaign. Most social media platforms indicate such a limit: Facebook 1000, TikTok 1000, LinkedIn 300, etc. There is one previous work \cite{fbnano} that showed this limit was not effectively imposed by Facebook at least until late 2020. Similarly, our research demonstrated that LinkedIn failed to impose its policy at least until July 2023. 

More generally, an important conclusion of our research is that advertising platforms should limit the amount of non-PII items that can be combined to define an audience. This practice will reduce the probability of uniquely identifying a user, reducing the feasibility of running nanotargeting ad campaigns.

\subsection{Dataset limitations}

We are aware that our dataset does not represent a random sample of the whole LinkedIn network. The 3352 users in our dataset are a small sample compared to the 970M users that are accessible through the LinkedIn advertising platform. To overcome this limitation we used three different sampling methods to show that the number of skills (plus location) required to uniquely identify a user on LinkedIn is low enough in all the cases to successfully run nanotargeting campaigns independently of the selected sampling method. Our main goal was to demonstrate that it was feasible to systematically nanotarget users with ads on LinkedIn using a combination of non-PII data items.

Our dataset has been used to obtain actual combinations of professional skills from active LinkedIn profiles. Those combinations have been later used in the LinkedIn Ads Platform to retrieve audience size values. Therefore, the audience size used in our methodology refers to the whole LinkedIn population and not only to the 3352 users in our sample. That information has been used to obtain the $N_P$ metric, employed as a reference in our nanotargeting campaigns. The outcome of our proof-of-concept experiment suggests that the $N_P$ metric values extracted from our model are reasonable, as shown in Table \ref{table:expected_probs_poc}. The expected number of successful campaigns based on the model results is very close to the successful campaigns in the experiment.

In summary, while we acknowledge that using a larger and more representative dataset had been better, the results suggest that, for the purpose of our paper, the collected dataset is good enough.
\section{Related work}
\label{sec:related_work}

In this section, we give an overview of the relevant literature related to this paper. We divide our literature revision into works addressing users' uniqueness and works focusing on nanotargeting. 

\subsection{Uniqueness}

We can find previous studies that have addressed user uniqueness in the context of social media and advertising \cite{fbnano} \cite{likes} \cite{twitter}, recommendation systems \cite{netflixdeanon}, mobile networks \cite{de2013unique}, credit card purchases \cite{de2015unique}, or even hidden services \cite{tor}. Most of the research in this area focuses on demonstrating how users' identities can be deanonymized using information that, in principle, may not be considered PII. 

A seminal work in this area demonstrated that combining gender, zip code, and birth date was enough to deanonymize the identity of 87\% of the 248M citizens in the 1990 US census \cite{sweeney2000simple}. This research was replicated for the 2000 US census, which included 281M citizens, and using the same combination of items, the percentage of identified users declined to 63\% \cite{Golle:2006:RUS:1179601.1179615}.  This means that 2 out of 3 users could be identified in a user base including 281M users combining only three (non-PII) data points: gender, zip code, and birth date. In this line, a recent work proposed a regenerative model to re-identify users based on demographic attributes and estimated that 99.98\% of Americans would be correctly re-identified in any dataset using 15 demographic attributes \cite{Nature_unique}.
Aligned to the previous result, we find two works that report the location and time associated with four mobile phone calls \cite{de2013unique} or four credit card purchases \cite{de2015unique} allow identifying a user in a database including $~$1M users with a probability $\geq90\%$. Again the conclusion is that 4 (non-PII) data points uniquely identify a user among 1M users. In another well-known work \cite{Netflix_Reidentification}, the authors managed to deanonymize the identity of some users embedded in the supposedly anonymized Netflix Prize dataset (480k users) \cite{Netflix_prize} using film rating entries from IMDB database. The authors proved that 8 movie ratings, along with an approximate date when they were completed, lead to a user identification probability of 99\%. 

We can also find a recent work that aims to identify users in hidden services, where anonymity is quite important \cite{tor}. Authors propose using publicly available Bitcoin payment data in those hidden services to unveil the user identity. The authors conducted a real experiment where they could link 125 unique users to 20 hidden services such as The Pirate Bay \cite{piratebay} and Silk Road \cite{silkroad}. 

Recent work also implements attribute inference attacks using publicly available non-PII data on players of the popular game Dota2\cite{Dota2AIA}. In that setting, an attacker can unveil private information from a user by exploiting publicly available information in the user profile. Although this work is not intended to make a user unique, it relates to our research since it exploits publicly available non-PII attributes to disclose a potential privacy issue.

More related to our study, some works look into user uniqueness in social media platforms. In \cite{likes}, the authors use fan page likes of Facebook users to deanonymize them. In \cite{twitter}, users' web browsing history is used to reveal the identity of those users on Twitter. Finally, in \cite{fbnano}, we demonstrate that 4 rare interests or 22 random interests Facebook has assigned a user for advertising purposes make a user unique with a probability of 90\% among almost 3 billion users. 

This paper extends the literature by showing how a novel piece of non-PII data, such as professional skills, can efficiently re-identify a user. Our results demonstrate that 6 data items (location and 5 skills) make a user unique on LinkedIn with a probability of 75\% among a user base of $\sim$970M users. However, there are two differential aspects in the case of LinkedIn compared to most studies in the literature: (i) the skills and location are information that users report in their LinkedIn profile, and it is publicly available. Contrarily, many of the previous works rely on information that is not easy to obtain, such as the browsing history of the user, record of phone calls or credit card purchases, list of ad preferences assigned by Facebook to a user, etc. (ii) The skills and location information was actionable to reach the users with tailored ads before LinkedIn fixed it. All the works in the literature but one, i.e., the one using FB ad preferences, do not address how the information used to identify a user can be activated to reach them.

\subsection{Nanotargeting}

The concept of nanotargeting is not new. Most of us are susceptible to receiving nanotargeting advertising through email, sms, or postal mail. Performing nanotargeting campaigns based on PII information such as email, mobile phone, or postal address is something trivial that has been implemented for many years. 

Also, some social media platforms, like Facebook, Twitter, LinkedIn, etc., offer advertisers the possibility of creating the so-called Custom Audiences or Matched Audiences to launch ad campaigns. A standard process to create a custom audience starts with the advertiser providing to the social media platform a file that includes PII information, such as email addresses or phone numbers, of the users the advertiser aims to target. The social media platform matches the provided PII information with its internal records to identify the user registered in the platform. The obtained list is the custom audience. From that moment, the advertiser can use that custom audience to deliver ads to the users included in the custom audience.

We can find real examples that have exploited the custom audience feature to run nanotargeting campaigns. In \cite{youtube}, one of the co-founders of Hawkers describes how they exploited nanotargeting based on custom audiences to target celebrities. He recognizes that they exploited Twitter's Custom Audiences \cite{twittercustomaudiences} to show an ad exclusively to Cristiano Ronaldo. Even though the size of custom audiences needs to be at least 1000 users, they included Cristiano Ronaldo using publicly available data in a list where he was the only man. After that, they filtered that audience to launch the campaign only to men in that custom audience (Twitter's implementation at the moment was limiting the size of the initial list, but the actual size of the targeted audience after applying additional filters was not checked). In other words,  they managed to nanotarget Cristiano Ronaldo. Their final goal was to make him aware of the brand and to approach him to arrange potential collaborations later, which is a clear example of user manipulation or influence using nanotargeting.

We can find other similar examples that have exploited custom audiences in various manners to nanotarget individuals \cite{prank, harf_2017,  tim_shipman_2018, hawkins_2019, faddoul2019sniper, haskins_2018,faizullabhoy2018facebooks,venkatadriPIIfb}. However, it is worth noting that using custom audiences to run nanotargeting campaigns fundamentally differs from our work since custom audiences are based on PII.

We can find a few examples in the literature that use non-PII attributes to implement nanotargeting campaigns. Dave Kerpen \cite{kerpen} conducted an experiment launching a campaign directed explicitly to his wife with the following parameters: 31-year-old married female, employees of Likeable Media living in New York City. The advert’s target audience was one user among hundreds of millions of users on Facebook at that time. The advert reached his wife, and only her, and she conducted the same experiment, also reaching Dave and only him with a response ad. In an academic work from 2010 \cite{korolova}, the authors accidentally rely on nanotargeting to infer additional (unknown) personal data from the targeted user.  This work describes a technique based on using fine-grained narrow audiences and campaign performance reports. The authors leave the unknown item they want to reveal as a free parameter (e.g., age). They later used multiple attributes retrieved from the Facebook profile of the user and ran multiple nanotargeting campaigns, each of them including a different value for the item they wanted to reveal. If all the campaigns report that they have reached 0 users but one of them indicates that it reached one user, the value of the unknown item is the one used in that campaign. 

Our work provides a much more comprehensive vision regarding the nanotargeting issue than the two cited works. First, we provide a theoretical analysis that explains the problem's dimension and the success probability of nanotargeting campaigns. Second, we run a proof of concept experiment that proves that nanotargeting could be systematically implemented within the LinkedIn advertising platform before they fixed it.

Only one previous work in the literature approaches nanotargeting using non-PII information systematically \cite{fbnano}. In this work, we analyzed how many ad preferences (i.e., interests FB assigns to users based on their activity to deliver them relevant ads) make a user unique on Facebook and whether that result can be used to activate a nanotargeting campaign. The conclusion is that 4 rare ad preferences or 22 random ad preferences are enough to make a user unique on Facebook with a probability of 90\%. In addition, this work demonstrates that running nanotargeting campaigns on Facebook is feasible based on users' ad preferences. 

The most important difference concerning this work is that ad preferences are not public information, and getting access to them is not trivial. We had to implement a web browser extension and achieve thousands of installations to obtain the dataset they used in their research. In other words, the potential nanotargeting attack described in that work is limited to skilled users capable of obtaining or inferring the ad preferences list of the individual they are targeting. In contrast, this work aims to validate the feasibility of running nanotargeting campaigns at scale using non-PII items anyone can access. That is why we focused on the skills and locations within LinkedIn. Our work does not only show running nanotargeting campaigns was feasible, but it also demonstrates it was plausible for low-skill attackers willing to implement it on LinkedIn before the fix.

\section{Ethics and legal considerations}
\label{sec:ethics}

Our research aims to reduce the privacy risks for users due to nanotargeting and to create awareness about how personal data, as defined in the General Data Protection Regulation (GDPR) \cite{gdpr} in the European Union, is different from Personal Identifiable Information (PII).

We believe this work is a relevant contribution in the context of the GDPR (and other advanced data protection regulations) as it provides a concrete example extending the vision of what should be considered personal data.

We expected that the combination of location and some professional skills would lead to a high uniqueness probability on LinkedIn. Hence, we took a conservative approach and considered we were managing personal data in our study, and thus, it was subject to the GDPR. We consulted with our institution's Data Protection Officer (DPO), based in the EU, to ensure compliance with the GDPR. The DPO confirmed that our research has a clear public interest as it aims to improve user privacy and helps to clarify whether the GDPR applies to specific combinations of data items. Therefore, the DPO confirmed that the legal basis supporting our research is the public interest, one of the legal bases exposed in the GDPR to allow personal data processing.

The only potential unique identifier we could have stored in our dataset was the URL used to access the LinkedIn profile. To protect user privacy, we replaced each profile's unique identifying URL with a random identifier to prevent anyone from potentially identifying individual users based on the information stored in our dataset. Following the instructions of our DPO, we implemented several security measures to minimize unauthorized access to our dataset. We kept the dataset in a server behind our institution's firewall and a second self-configured firewall. The server is only accessible from a device connected to our institution's physical network or VPN. Server access requires having an account and password on the server. Finally, the dataset was encrypted, and only the paper's authors had the credentials to access the information. We adopted these security measures to safeguard the data from unauthorized access and comply with the requirements of the GDPR.

In summary, this research is closely linked to ethical principles and aims to reduce the privacy risks of users on LinkedIn and enhance the application of the GDPR. Furthermore, we have ensured compliance with the GDPR by following the instructions of our institution's DPO, who reviewed and approved this research work.

\section{Conclusions}
\label{sec:conclusions}

This work contributes to the body of literature demonstrating that a few non-PII data items are enough to uniquely identify a user among a user base of tens or hundreds of millions of users.  Our work shows that online privacy is very vulnerable, and anonymizing unique identifiers is not enough to hide users' identities and protect them from being targeted.

The main contribution of our work is that we have shown for the first time that publicly available data could be exploited by third parties to potentially target hundreds of millions of users with hyper-personalized ads individually. The literature refers to this practice as nanotargeting. Nanotargeting may expose users to privacy risks derived from malicious activities such as malvertising, manipulation, or blackmail. In our opinion, our work unveiled a huge privacy gap that had to be urgently covered. Fortunately, despite LinkedIn ignoring our communication regarding the vulnerability we found to nanotarget users through their advertising platform, LinkedIn has fixed it and as of November 2023, we were unable to implement nanotargeting campaigns in LinkedIn. This measure eliminates the risk of nanotargeting on this platform. Now, ads can still be delivered to the targeted user, but you need to define an audience including at least 299 other users. Therefore, the ad may not exclusively reach the targeted user. We note this still represents a risk for the targeted user

We are extremely happy that our research may have been useful in improving the security and privacy of hundreds of millions of users since our lower bound estimations revealed that at least a quarter billion users on LinkedIn were exposed to nanotargeting ad campaigns with a high probability of success. It is important to show success stories around academic privacy research that encourage the community to unveil privacy vulnerabilities knowing that they may have a real impact through their remediation. From the research point of view, the bug fix implies that our study cannot be reproduced.

Despite LinkedIn fixing the discovered vulnerabilities there may be other advertising platforms on the Internet that may still be exploited to run nanotargeting campaigns based on publicly available non-PII.  We propose two immediate actions to mitigate the undesirable effects unveiled by our research.

First, any advertising platform should impose effective countermeasures that preclude advertisers from running nanotargeting campaigns based on combinations of non-PII attributes. The solution is extremely simple. As LinkedIn is finally doing, advertising platforms have to effectively impose policies that impede advertisers running campaigns targeting less than a few hundred or a few thousand users. The higher the threshold the better in terms of users' privacy.
%We struggle to understand why LinkedIn was failing to implement that policy since, in our opinion, they did not have any clear incentive for not doing it. %Unfortunately, the reaction of LinkedIn to our responsible disclosure procedure is quite disappointing since they do not consider the bug unveiled in our work as a vulnerability. %$(i)$ LinkedIn is likely not meeting the GDPR requirements since, to the best of our knowledge, users have not given explicit permission to be nanotargeted based on non-PII attributes. $(ii)$  Nanotargeting campaigns generate negligible revenue compared with wider ad campaigns targeting hundreds or thousands of users. The key question is whether the lack of implementation of the advertised policy is simply an involuntary mistake or it is a conscious decision. We believe the actual reason is the former due to the referred lack of incentives. 
 
Second, our work discusses the practical limitations of the current definition of personal data in the GDPR to assess whether a combination of non-PII elements should be considered personal data. This generates uncertainties in the application of the GDPR since demonstrating whether a combination of certain non-PII items allows uniquely identifying a user is a complex task, even for companies and regulators. Data protection authorities should work with the research community to elaborate a guide of good practices in managing non-PII data. This guide should define a clear ground for companies regarding when they should consider combinations of non-PII as personal data. At the same time, that guide may also help citizens to better identify potentially risky situations for their privacy. 

\begin{acks}
This work has been partially funded by the following projects:  the project TESTABLE (Grant 101019206) funded by European Union’s Horizon 2020 programme; the project AUDINT (Grant TED2021-132076B-I00) funded by the MCIN/AEI/ 10.13039/501100011033 and the EU NextGeneration/PRTR funds; the A2 PRIVCOMP funded by the Ministerio de Asuntos Económicos y Transformación Digital and the European Union-NextGenerationEU.
\end{acks}

{\footnotesize \bibliographystyle{ACM-Reference-Format}
\bibliography{docs/sample.bib}}

%%% -*-BibTeX-*-
%%% Do NOT edit. File created by BibTeX with style
%%% ACM-Reference-Format-Journals [18-Jan-2012].

\begin{thebibliography}{54}

%%% ====================================================================
%%% NOTE TO THE USER: you can override these defaults by providing
%%% customized versions of any of these macros before the \bibliography
%%% command.  Each of them MUST provide its own final punctuation,
%%% except for \shownote{}, \showDOI{}, and \showURL{}.  The latter two
%%% do not use final punctuation, in order to avoid confusing it with
%%% the Web address.
%%%
%%% To suppress output of a particular field, define its macro to expand
%%% to an empty string, or better, \unskip, like this:
%%%
%%% \newcommand{\showDOI}[1]{\unskip}   % LaTeX syntax
%%%
%%% \def \showDOI #1{\unskip}           % plain TeX syntax
%%%
%%% ====================================================================

\ifx \showCODEN    \undefined \def \showCODEN     #1{\unskip}     \fi
\ifx \showDOI      \undefined \def \showDOI       #1{#1}\fi
\ifx \showISBNx    \undefined \def \showISBNx     #1{\unskip}     \fi
\ifx \showISBNxiii \undefined \def \showISBNxiii  #1{\unskip}     \fi
\ifx \showISSN     \undefined \def \showISSN      #1{\unskip}     \fi
\ifx \showLCCN     \undefined \def \showLCCN      #1{\unskip}     \fi
\ifx \shownote     \undefined \def \shownote      #1{#1}          \fi
\ifx \showarticletitle \undefined \def \showarticletitle #1{#1}   \fi
\ifx \showURL      \undefined \def \showURL       {\relax}        \fi
% The following commands are used for tagged output and should be
% invisible to TeX
\providecommand\bibfield[2]{#2}
\providecommand\bibinfo[2]{#2}
\providecommand\natexlab[1]{#1}
\providecommand\showeprint[2][]{arXiv:#2}

\bibitem[Arrate et~al\mbox{.}(2020)]%
        {Malvertising_Aritz}
\bibfield{author}{\bibinfo{person}{Aritz Arrate}, \bibinfo{person}{José González-Cabañas}, \bibinfo{person}{{\'{A}}ngel Cuevas}, {and} \bibinfo{person}{Rubén Cuevas}.} \bibinfo{year}{2020}\natexlab{}.
\newblock \showarticletitle{Malvertising in Facebook: Analysis, Quantification and Solution}.
\newblock \bibinfo{journal}{\emph{Electronics}} \bibinfo{volume}{9}, \bibinfo{number}{8} (\bibinfo{year}{2020}).
\newblock
\showISSN{2079-9292}
\urldef\tempurl%
\url{https://doi.org/10.3390/electronics9081332}
\showDOI{\tempurl}


\bibitem[Barratt(2012)]%
        {silkroad}
\bibfield{author}{\bibinfo{person}{Monica~J. Barratt}.} \bibinfo{year}{2012}\natexlab{}.
\newblock \showarticletitle{SILK ROAD: EBAY FOR DRUGS}.
\newblock \bibinfo{journal}{\emph{Addiction}} \bibinfo{volume}{107}, \bibinfo{number}{3} (\bibinfo{year}{2012}), \bibinfo{pages}{683--683}.
\newblock
\showISSN{0965-2140}
\urldef\tempurl%
\url{https://doi.org/10.1111/j.1360-0443.2011.03709.x}
\showDOI{\tempurl}


\bibitem[Bennett and Lanning(2007)]%
        {Netflix_prize}
\bibfield{author}{\bibinfo{person}{J. Bennett} {and} \bibinfo{person}{S. Lanning}.} \bibinfo{year}{2007}\natexlab{}.
\newblock \showarticletitle{The Netflix Prize}. In \bibinfo{booktitle}{\emph{Proceedings of the KDD Cup Workshop 2007}}. \bibinfo{publisher}{ACM}, \bibinfo{address}{New York}, \bibinfo{pages}{3--6}.
\newblock
\urldef\tempurl%
\url{http://www.cs.uic.edu/~liub/KDD-cup-2007/NetflixPrize-description.pdf}
\showURL{%
\tempurl}


\bibitem[Bundy and Wallen(1984)]%
        {BFS}
\bibfield{author}{\bibinfo{person}{Alan Bundy} {and} \bibinfo{person}{Lincoln Wallen}.} \bibinfo{year}{1984}\natexlab{}.
\newblock \bibinfo{booktitle}{\emph{Breadth-First Search}}.
\newblock \bibinfo{publisher}{Springer Berlin Heidelberg}, \bibinfo{address}{Berlin, Heidelberg}, \bibinfo{pages}{13--13}.
\newblock
\showISBNx{978-3-642-96868-6}
\urldef\tempurl%
\url{https://doi.org/10.1007/978-3-642-96868-6\_25}
\showDOI{\tempurl}


\bibitem[Cesario et~al\mbox{.}(2008)]%
        {pshyc_persuasion1}
\bibfield{author}{\bibinfo{person}{Joseph Cesario}, \bibinfo{person}{E.~Tory Higgins}, {and} \bibinfo{person}{Abigail~A. Scholer}.} \bibinfo{year}{2008}\natexlab{}.
\newblock \showarticletitle{{Regulatory Fit and Persuasion: Basic Principles and Remaining Questions}}.
\newblock \bibinfo{journal}{\emph{Social and Personality Psychology Compass}} \bibinfo{volume}{2}, \bibinfo{number}{1} (\bibinfo{year}{2008}), \bibinfo{pages}{444--463}.
\newblock
\urldef\tempurl%
\url{https://doi.org/10.1111/j.1751-9004.2007.00055.x}
\showDOI{\tempurl}
\showeprint{https://onlinelibrary.wiley.com/doi/pdf/10.1111/j.1751-9004.2007.00055.x}


\bibitem[Chen et~al\mbox{.}(2019)]%
        {Malvertising_WWW}
\bibfield{author}{\bibinfo{person}{Gong Chen}, \bibinfo{person}{Wei Meng}, {and} \bibinfo{person}{John Copeland}.} \bibinfo{year}{2019}\natexlab{}.
\newblock \showarticletitle{Revisiting Mobile Advertising Threats with MAdLife}. In \bibinfo{booktitle}{\emph{The World Wide Web Conference}} (San Francisco, CA, USA) \emph{(\bibinfo{series}{WWW '19})}. \bibinfo{publisher}{ACM}, \bibinfo{address}{New York, NY, USA}, \bibinfo{pages}{207--217}.
\newblock
\showISBNx{978-1-4503-6674-8}
\urldef\tempurl%
\url{https://doi.org/10.1145/3308558.3313549}
\showDOI{\tempurl}


\bibitem[Cho et~al\mbox{.}(2016)]%
        {Malvertising_mobile1}
\bibfield{author}{\bibinfo{person}{Geumhwan Cho}, \bibinfo{person}{Junsung Cho}, \bibinfo{person}{Youngbae Song}, \bibinfo{person}{Donghyun Choi}, {and} \bibinfo{person}{Hyoungshick Kim}.} \bibinfo{year}{2016}\natexlab{}.
\newblock \showarticletitle{Combating online fraud attacks in mobile-based advertising}.
\newblock \bibinfo{journal}{\emph{EURASIP Journal on Information Security}} \bibinfo{volume}{2016}, \bibinfo{number}{1} (\bibinfo{date}{04 Jan} \bibinfo{year}{2016}), \bibinfo{pages}{2}.
\newblock
\showISSN{1687-417X}
\urldef\tempurl%
\url{https://doi.org/10.1186/s13635-015-0027-7}
\showDOI{\tempurl}


\bibitem[Corporation(2022)]%
        {ldvisibility}
\bibfield{author}{\bibinfo{person}{LinkedIn Corporation}.} \bibinfo{year}{2022}\natexlab{}.
\newblock \bibinfo{booktitle}{\emph{What People Can See on Your Profile}}.
\newblock


\bibitem[Corporation(2023)]%
        {ldstats}
\bibfield{author}{\bibinfo{person}{LinkedIn Corporation}.} \bibinfo{year}{2023}\natexlab{}.
\newblock \bibinfo{booktitle}{\emph{About Us: Statistics}}.
\newblock


\bibitem[de~Montjoye et~al\mbox{.}(2013)]%
        {de2013unique}
\bibfield{author}{\bibinfo{person}{Yves-Alexandre de Montjoye}, \bibinfo{person}{César~A Hidalgo}, \bibinfo{person}{Michel Verleysen}, {and} \bibinfo{person}{Vincent~D Blondel}.} \bibinfo{year}{2013}\natexlab{}.
\newblock \showarticletitle{{Unique in the Crowd: The privacy bounds of human mobility}}.
\newblock \bibinfo{journal}{\emph{Scientific reports}} \bibinfo{volume}{3}, \bibinfo{number}{1} (\bibinfo{year}{2013}), \bibinfo{pages}{1376}.
\newblock
\showISSN{2045-2322}
\urldef\tempurl%
\url{https://www.nature.com/articles/srep01376}
\showURL{%
\tempurl}


\bibitem[De~Montjoye et~al\mbox{.}(2015)]%
        {de2015unique}
\bibfield{author}{\bibinfo{person}{Yves-Alexandre De~Montjoye}, \bibinfo{person}{Laura Radaelli}, \bibinfo{person}{Vivek~Kumar Singh}, {et~al\mbox{.}}} \bibinfo{year}{2015}\natexlab{}.
\newblock \showarticletitle{{Unique in the shopping mall: On the reidentifiability of credit card metadata}}.
\newblock \bibinfo{journal}{\emph{Science}} \bibinfo{volume}{347}, \bibinfo{number}{6221} (\bibinfo{year}{2015}), \bibinfo{pages}{536--539}.
\newblock
\urldef\tempurl%
\url{https://science.sciencemag.org/content/347/6221/536}
\showURL{%
\tempurl}


\bibitem[Dubois et~al\mbox{.}(2016)]%
        {pshyc_persuasion5}
\bibfield{author}{\bibinfo{person}{David Dubois}, \bibinfo{person}{Derek~D. Rucker}, {and} \bibinfo{person}{Adam~D. Galinsky}.} \bibinfo{year}{2016}\natexlab{}.
\newblock \showarticletitle{{Dynamics of Communicator and Audience Power: The Persuasiveness of Competence versus Warmth}}.
\newblock \bibinfo{journal}{\emph{Journal of Consumer Research}} \bibinfo{volume}{43}, \bibinfo{number}{1} (\bibinfo{date}{Feb.} \bibinfo{year}{2016}), \bibinfo{pages}{68--85}.
\newblock
\showISSN{0093-5301}
\urldef\tempurl%
\url{https://doi.org/10.1093/jcr/ucw006}
\showDOI{\tempurl}
\showeprint{https://academic.oup.com/jcr/article-pdf/43/1/68/7049938/ucw006.pdf}


\bibitem[EU(2016)]%
        {gdpr}
\bibfield{author}{\bibinfo{person}{EU}.} \bibinfo{year}{2016}\natexlab{}.
\newblock \bibinfo{title}{{Regulation (EU) 2016/679 of the European Parliament and of the Council of 27 April 2016 on the protection of natural persons with regard to the processing of personal data and on the free movement of such data, and repealing Directive 95/46/EC (General Data Protection Regulation)}}.
\newblock \bibinfo{howpublished}{European Union}.
\newblock
\urldef\tempurl%
\url{http://eur-lex.europa.eu/eli/reg/2016/679/oj}
\showURL{%
\tempurl}
\newblock
\shownote{accessed on 21 September, 2021}.


\bibitem[Faddoul et~al\mbox{.}(2019)]%
        {faddoul2019sniper}
\bibfield{author}{\bibinfo{person}{Marc Faddoul}, \bibinfo{person}{Rohan Kapuria}, {and} \bibinfo{person}{Lily Lin}.} \bibinfo{year}{2019}\natexlab{}.
\newblock \showarticletitle{SNIPER AD TARGETING}.
\newblock \bibinfo{journal}{\emph{Berkeley School of Information}} (\bibinfo{date}{May} \bibinfo{year}{2019}).
\newblock
\urldef\tempurl%
\url{https://www.ischool.berkeley.edu/projects/2019/sniper-ad-targeting}
\showURL{%
\tempurl}


\bibitem[Faizullabhoy and Korolova(2018)]%
        {faizullabhoy2018facebooks}
\bibfield{author}{\bibinfo{person}{Irfan Faizullabhoy} {and} \bibinfo{person}{Aleksandra Korolova}.} \bibinfo{year}{2018}\natexlab{}.
\newblock \bibinfo{title}{Facebook's Advertising Platform: New Attack Vectors and the Need for Interventions}.
\newblock
\newblock
\showeprint[arxiv]{1803.10099}~[cs.CY]


\bibitem[Ford et~al\mbox{.}(2009)]%
        {flashmalvert}
\bibfield{author}{\bibinfo{person}{Sean Ford}, \bibinfo{person}{Marco Cova}, \bibinfo{person}{Christopher Kruegel}, {and} \bibinfo{person}{Giovanni Vigna}.} \bibinfo{year}{2009}\natexlab{}.
\newblock \showarticletitle{Analyzing and Detecting Malicious Flash Advertisements}. In \bibinfo{booktitle}{\emph{2009 Annual Computer Security Applications Conference}}. \bibinfo{pages}{363--372}.
\newblock
\urldef\tempurl%
\url{https://doi.org/10.1109/ACSAC.2009.41}
\showDOI{\tempurl}


\bibitem[Golle(2006)]%
        {Golle:2006:RUS:1179601.1179615}
\bibfield{author}{\bibinfo{person}{Philippe Golle}.} \bibinfo{year}{2006}\natexlab{}.
\newblock \showarticletitle{{Revisiting the Uniqueness of Simple Demographics in the US Population}}. In \bibinfo{booktitle}{\emph{Proceedings of the 5th ACM Workshop on Privacy in Electronic Society}} (Alexandria, Virginia, USA) \emph{(\bibinfo{series}{WPES '06})}. \bibinfo{publisher}{ACM}, \bibinfo{address}{New York, NY, USA}, \bibinfo{pages}{77--80}.
\newblock
\showISBNx{1-59593-556-8}
\urldef\tempurl%
\url{https://doi.org/10.1145/1179601.1179615}
\showDOI{\tempurl}


\bibitem[Gonz\'{a}lez-Caba\~{n}as et~al\mbox{.}(2021)]%
        {fbnano}
\bibfield{author}{\bibinfo{person}{Jos\'{e} Gonz\'{a}lez-Caba\~{n}as}, \bibinfo{person}{\'{A}ngel Cuevas}, \bibinfo{person}{Rub\'{e}n Cuevas}, \bibinfo{person}{Juan L\'{o}pez-Fern\'{a}ndez}, {and} \bibinfo{person}{David Garc\'{\i}a}.} \bibinfo{year}{2021}\natexlab{}.
\newblock \showarticletitle{Unique on Facebook: Formulation and Evidence of (Nano)Targeting Individual Users with Non-PII Data}. In \bibinfo{booktitle}{\emph{Proceedings of the 21st ACM Internet Measurement Conference}} (Virtual Event) \emph{(\bibinfo{series}{IMC '21})}. \bibinfo{publisher}{Association for Computing Machinery}, \bibinfo{address}{New York, NY, USA}, \bibinfo{pages}{464–479}.
\newblock
\showISBNx{9781450391290}
\urldef\tempurl%
\url{https://doi.org/10.1145/3487552.3487861}
\showDOI{\tempurl}


\bibitem["HackerOne"(2023)]%
        {ldhackerone}
\bibfield{author}{\bibinfo{person}{"HackerOne"}.} \bibinfo{year}{2023}\natexlab{}.
\newblock \bibinfo{booktitle}{\emph{LinkedIn's Bug Bounty Program}}.
\newblock
\urldef\tempurl%
\url{https://hackerone.com/linkedin}
\showURL{%
\tempurl}


\bibitem[Harf(2017)]%
        {harf_2017}
\bibfield{author}{\bibinfo{person}{Michael Harf}.} \bibinfo{year}{2017}\natexlab{}.
\newblock \bibinfo{title}{{Sniper Targeting on Facebook: How to Target ONE specific person with super targeted ads}}.
\newblock \bibinfo{howpublished}{Medium}.
\newblock
\urldef\tempurl%
\url{https://medium.com/@MichaelH\_3009/sniper-targeting-on-facebook-how-to-target-one-specific-person-with-super-targeted-ads-515ba6e068f6}
\showURL{%
\tempurl}
\newblock
\shownote{accessed on 29 January, 2023}.


\bibitem[Haskins(2018)]%
        {haskins_2018}
\bibfield{author}{\bibinfo{person}{Caroline Haskins}.} \bibinfo{year}{2018}\natexlab{}.
\newblock \bibinfo{title}{Facebook ad micro-targeting can manipulate individual politicians}.
\newblock \bibinfo{howpublished}{The Outline}.
\newblock
\urldef\tempurl%
\url{https://theoutline.com/post/5411/facebook-ad-micro-targeting-can-manipulate-individual-politicians}
\showURL{%
\tempurl}
\newblock
\shownote{accessed on 29 January, 2023}.


\bibitem[Hawkins(2019)]%
        {hawkins_2019}
\bibfield{author}{\bibinfo{person}{Jonathan Hawkins}.} \bibinfo{year}{2019}\natexlab{}.
\newblock \bibinfo{title}{{Facebook Ads Sniper Method: How to Put Your Ad in front of ONE Specific Person}}.
\newblock \bibinfo{howpublished}{Jonathan Hawkins}.
\newblock
\urldef\tempurl%
\url{https://medium.com/@jonathanhawkinsofficial/facebook-ads-sniper-method-how-to-put-your-ad-in-front-of-one-specific-person-7e96a2e3984c}
\showURL{%
\tempurl}
\newblock
\shownote{accessed on 29 January, 2023}.


\bibitem[Hirsh et~al\mbox{.}(2012)]%
        {pshyc_persuasion4}
\bibfield{author}{\bibinfo{person}{Jacob~B. Hirsh}, \bibinfo{person}{Sonia~K. Kang}, {and} \bibinfo{person}{Galen~V. Bodenhausen}.} \bibinfo{year}{2012}\natexlab{}.
\newblock \showarticletitle{{Personalized Persuasion: Tailoring Persuasive Appeals to Recipients’ Personality Traits}}.
\newblock \bibinfo{journal}{\emph{Psychological Science}} \bibinfo{volume}{23}, \bibinfo{number}{6} (\bibinfo{year}{2012}), \bibinfo{pages}{578--581}.
\newblock
\urldef\tempurl%
\url{https://doi.org/10.1177/0956797611436349}
\showDOI{\tempurl}
\showeprint{https://doi.org/10.1177/0956797611436349}
\newblock
\shownote{PMID: 22547658}.


\bibitem[Huang et~al\mbox{.}(2018)]%
        {Malvertising_Bayesian}
\bibfield{author}{\bibinfo{person}{Chin-Tser Huang}, \bibinfo{person}{Muhammad Sakib}, \bibinfo{person}{Charles Kamhoua}, \bibinfo{person}{Kevin Kwiat}, {and} \bibinfo{person}{Laurent Njilla}.} \bibinfo{year}{2018}\natexlab{}.
\newblock \showarticletitle{A Bayesian Game Theoretic Approach for Inspecting Web-based Malvertising}.
\newblock \bibinfo{journal}{\emph{IEEE Transactions on Dependable and Secure Computing}}  \bibinfo{volume}{PP} (\bibinfo{date}{08} \bibinfo{year}{2018}), \bibinfo{pages}{1--1}.
\newblock
\urldef\tempurl%
\url{https://doi.org/10.1109/tdsc.2018.2866821}
\showDOI{\tempurl}


\bibitem[Jawaheri et~al\mbox{.}(2020)]%
        {tor}
\bibfield{author}{\bibinfo{person}{Husam~Al Jawaheri}, \bibinfo{person}{Mashael~Al Sabah}, \bibinfo{person}{Yazan Boshmaf}, {and} \bibinfo{person}{Aiman Erbad}.} \bibinfo{year}{2020}\natexlab{}.
\newblock \showarticletitle{Deanonymizing Tor hidden service users through Bitcoin transactions analysis}.
\newblock \bibinfo{howpublished}{\url{https://www.sciencedirect.com/science/article/pii/S0167404818309908}}.
\newblock \bibinfo{journal}{\emph{Computers \& Security}}  \bibinfo{volume}{89} (\bibinfo{year}{2020}), \bibinfo{pages}{101684}.
\newblock
\showISSN{0167-4048}
\urldef\tempurl%
\url{https://doi.org/10.1016/j.cose.2019.101684}
\showDOI{\tempurl}


\bibitem[Kerpen(2011)]%
        {kerpen}
\bibfield{author}{\bibinfo{person}{Dave Kerpen}.} \bibinfo{year}{2011}\natexlab{}.
\newblock \bibinfo{booktitle}{\emph{Likeable Social Media: How to Delight Your Customers, Create an Irresistible Brand, and Be Generally Amazing on Facebook (\& Other Social Networks)}}.
\newblock \bibinfo{publisher}{McGraw Hill-Ascent Audio}.
\newblock
\showISBNx{0071804552}


\bibitem[Korolova(2010)]%
        {korolova}
\bibfield{author}{\bibinfo{person}{Aleksandra Korolova}.} \bibinfo{year}{2010}\natexlab{}.
\newblock \showarticletitle{Privacy Violations Using Microtargeted Ads: A Case Study}. In \bibinfo{booktitle}{\emph{2010 IEEE International Conference on Data Mining Workshops}}. \bibinfo{pages}{474--482}.
\newblock
\urldef\tempurl%
\url{https://doi.org/10.1109/ICDMW.2010.137}
\showDOI{\tempurl}


\bibitem[LinkedIn(2023a)]%
        {ldcampaignmanager}
\bibfield{author}{\bibinfo{person}{LinkedIn}.} \bibinfo{year}{2023}\natexlab{a}.
\newblock \bibinfo{booktitle}{\emph{LinkedIn Marketing Solutions}}.
\newblock
\urldef\tempurl%
\url{https://business.linkedin.com/es-es/marketing-solutions}
\showURL{%
\tempurl}


\bibitem[LinkedIn(2023b)]%
        {ldsecurity}
\bibfield{author}{\bibinfo{person}{LinkedIn}.} \bibinfo{year}{2023}\natexlab{b}.
\newblock \bibinfo{booktitle}{\emph{LinkedIn Security}}.
\newblock
\urldef\tempurl%
\url{https://security.linkedin.com/}
\showURL{%
\tempurl}


\bibitem[LinkedIn(2023c)]%
        {ldhelpaudience}
\bibfield{author}{\bibinfo{person}{LinkedIn}.} \bibinfo{year}{2023}\natexlab{c}.
\newblock \bibinfo{booktitle}{\emph{Target Audience Size – Best Practices}}.
\newblock


\bibitem[Mansfield-Devine(2014)]%
        {Mansfield-Devine-malvert}
\bibfield{author}{\bibinfo{person}{Steve Mansfield-Devine}.} \bibinfo{year}{2014}\natexlab{}.
\newblock \showarticletitle{The dark side of advertising}.
\newblock \bibinfo{journal}{\emph{Computer Fraud \& Security}} \bibinfo{volume}{2014}, \bibinfo{number}{11} (\bibinfo{year}{2014}), \bibinfo{pages}{5--8}.
\newblock
\showISSN{1361-3723}
\urldef\tempurl%
\url{https://doi.org/10.1016/S1361-3723(14)70547-0}
\showDOI{\tempurl}


\bibitem[Matz et~al\mbox{.}(2017)]%
        {Kosinski_PNAS}
\bibfield{author}{\bibinfo{person}{S.~C. Matz}, \bibinfo{person}{M. Kosinski}, \bibinfo{person}{G. Nave}, {and} \bibinfo{person}{D.~J. Stillwell}.} \bibinfo{year}{2017}\natexlab{}.
\newblock \showarticletitle{Psychological targeting as an effective approach to digital mass persuasion}.
\newblock \bibinfo{journal}{\emph{Proceedings of the National Academy of Sciences}} \bibinfo{volume}{114}, \bibinfo{number}{48} (\bibinfo{year}{2017}), \bibinfo{pages}{12714--12719}.
\newblock
\showISSN{0027-8424}
\urldef\tempurl%
\url{https://doi.org/10.1073/pnas.1710966114}
\showDOI{\tempurl}


\bibitem[Moon(2002)]%
        {pshyc_persuasion3}
\bibfield{author}{\bibinfo{person}{Youngme Moon}.} \bibinfo{year}{2002}\natexlab{}.
\newblock \showarticletitle{{Personalization and Personality: Some Effects of Customizing Message Style Based on Consumer Personality}}.
\newblock \bibinfo{journal}{\emph{Journal of Consumer Psychology}} \bibinfo{volume}{12}, \bibinfo{number}{4} (\bibinfo{year}{2002}), \bibinfo{pages}{313--325}.
\newblock
\urldef\tempurl%
\url{https://doi.org/10.1016/S1057-7408(16)30083-3}
\showDOI{\tempurl}
\showeprint{https://onlinelibrary.wiley.com/doi/pdf/10.1016/S1057-7408\%2816\%2930083-3}


\bibitem[Moreno et~al\mbox{.}({[n.\,d.]})]%
        {youtube}
\bibfield{author}{\bibinfo{person}{David Moreno}, \bibinfo{person}{Alex Moreno}, \bibinfo{person}{Alex Benlloch}, {and} \bibinfo{person}{Bruno Casanovas}.} \bibinfo{year}{[n.\,d.]}\natexlab{}.
\newblock \bibinfo{booktitle}{\emph{De 0 a 100 millones, HAWKERS}}.
\newblock Youtube.
\newblock


\bibitem[{Mullock} et~al\mbox{.}(2010)]%
        {mullock}
\bibfield{author}{\bibinfo{person}{J. {Mullock}}, \bibinfo{person}{S. {Groom}}, \bibinfo{person}{}, {and} \bibinfo{person}{P. {Lee}}.} \bibinfo{year}{2010}\natexlab{}.
\newblock \bibinfo{title}{International online behavioural advertising survey 2010}.
\newblock \bibinfo{howpublished}{Osborne Clarke}.
\newblock


\bibitem[Narayanan and Shmatikov(2008a)]%
        {netflixdeanon}
\bibfield{author}{\bibinfo{person}{Arvind Narayanan} {and} \bibinfo{person}{Vitaly Shmatikov}.} \bibinfo{year}{2008}\natexlab{a}.
\newblock \showarticletitle{Robust De-anonymization of Large Sparse Datasets}. In \bibinfo{booktitle}{\emph{2008 IEEE Symposium on Security and Privacy (sp 2008)}}. \bibinfo{pages}{111--125}.
\newblock
\urldef\tempurl%
\url{https://doi.org/10.1109/SP.2008.33}
\showDOI{\tempurl}


\bibitem[Narayanan and Shmatikov(2008b)]%
        {Netflix_Reidentification}
\bibfield{author}{\bibinfo{person}{Arvind Narayanan} {and} \bibinfo{person}{Vitaly Shmatikov}.} \bibinfo{year}{2008}\natexlab{b}.
\newblock \showarticletitle{{Robust De-Anonymization of Large Sparse Datasets}}. In \bibinfo{booktitle}{\emph{Proceedings of the 2008 IEEE Symposium on Security and Privacy}} \emph{(\bibinfo{series}{SP '08})}. \bibinfo{publisher}{IEEE Computer Society}, \bibinfo{address}{USA}, \bibinfo{pages}{111–125}.
\newblock
\showISBNx{9780769531687}
\urldef\tempurl%
\url{https://doi.org/10.1109/SP.2008.33}
\showDOI{\tempurl}


\bibitem[Poornachandran et~al\mbox{.}(2017)]%
        {Malvertising_SVM}
\bibfield{author}{\bibinfo{person}{Prabaharan Poornachandran}, \bibinfo{person}{N. Balagopal}, \bibinfo{person}{Soumajit Pal}, \bibinfo{person}{Aravind Ashok}, \bibinfo{person}{Prem Sankar}, {and} \bibinfo{person}{Manu~R. Krishnan}.} \bibinfo{year}{2017}\natexlab{}.
\newblock \showarticletitle{Demalvertising: A Kernel Approach for Detecting Malwares in Advertising Networks}. In \bibinfo{booktitle}{\emph{Proceedings of the First International Conference on Intelligent Computing and Communication}}, \bibfield{editor}{\bibinfo{person}{Jyotsna~Kumar Mandal}, \bibinfo{person}{Suresh~Chandra Satapathy}, \bibinfo{person}{Manas~Kumar Sanyal}, {and} \bibinfo{person}{Vikrant Bhateja}} (Eds.). \bibinfo{publisher}{Springer Singapore}, \bibinfo{address}{Singapore}, \bibinfo{pages}{215--224}.
\newblock


\bibitem[Rastogi et~al\mbox{.}(2016)]%
        {Malvertising_NDSS}
\bibfield{author}{\bibinfo{person}{Vaibhav Rastogi}, \bibinfo{person}{Rui Shao}, \bibinfo{person}{Yan Chen}, \bibinfo{person}{Xiang Pan}, \bibinfo{person}{Shihong Zou}, {and} \bibinfo{person}{Ryan Riley}.} \bibinfo{year}{2016}\natexlab{}.
\newblock \showarticletitle{Are these Ads Safe: Detecting Hidden Attacks through the Mobile App-Web Interfaces}. In \bibinfo{booktitle}{\emph{NDSS}}.
\newblock
\urldef\tempurl%
\url{https://doi.org/10.14722/ndss.2016.23234}
\showDOI{\tempurl}


\bibitem[Rocher et~al\mbox{.}(2019)]%
        {Nature_unique}
\bibfield{author}{\bibinfo{person}{Luc Rocher}, \bibinfo{person}{Julien~M. Hendrickx}, {and} \bibinfo{person}{Yves-Alexandre de Montjoye}.} \bibinfo{year}{2019}\natexlab{}.
\newblock \showarticletitle{Estimating the success of re-identifications in incomplete datasets using generative models}.
\newblock \bibinfo{journal}{\emph{Nature Communications}} \bibinfo{volume}{10}, \bibinfo{number}{1} (\bibinfo{year}{2019}), \bibinfo{pages}{3069}.
\newblock
\showISBNx{2041-1723}
\urldef\tempurl%
\url{https://doi.org/10.1038/s41467-019-10933-3}
\showDOI{\tempurl}


\bibitem[Roth(2020)]%
        {ldtopvoices}
\bibfield{author}{\bibinfo{person}{Daniel Roth}.} \bibinfo{year}{2020}\natexlab{}.
\newblock \bibinfo{booktitle}{\emph{LinkedIn Top Voices 2020: Meet the professionals driving today’s business conversation.}}
\newblock


\bibitem[Rüdian et~al\mbox{.}(2018)]%
        {likes}
\bibfield{author}{\bibinfo{person}{S. Rüdian}, \bibinfo{person}{N. Pinkwart}, {and} \bibinfo{person}{Z. Liu}.} \bibinfo{year}{2018}\natexlab{}.
\newblock \showarticletitle{I know who you are: Deanonymization using Facebook likes}. In \bibinfo{booktitle}{\emph{Lecture Notes in Informatics (LNI), Proceedings - Series of the Gesellschaft fur Informatik (GI)}}, Vol.~\bibinfo{volume}{285}. \bibinfo{pages}{109--118}.
\newblock


\bibitem[Sakib and Huang(2015)]%
        {maladscollection}
\bibfield{author}{\bibinfo{person}{Muhammad~N. Sakib} {and} \bibinfo{person}{Chin-Tser Huang}.} \bibinfo{year}{2015}\natexlab{}.
\newblock \showarticletitle{Automated Collection and Analysis of Malware Disseminated via Online Advertising}. In \bibinfo{booktitle}{\emph{2015 IEEE Trustcom/BigDataSE/ISPA}}, Vol.~\bibinfo{volume}{1}. \bibinfo{pages}{1411--1416}.
\newblock
\urldef\tempurl%
\url{https://doi.org/10.1109/Trustcom.2015.539}
\showDOI{\tempurl}


\bibitem[Shapero(2021)]%
        {ldrevenue}
\bibfield{author}{\bibinfo{person}{Daniel Shapero}.} \bibinfo{year}{2021}\natexlab{}.
\newblock \bibinfo{booktitle}{\emph{LinkedIn hits \$10B in revenue by helping companies connect with talent and customers.}}
\newblock


\bibitem[Su et~al\mbox{.}(2017)]%
        {twitter}
\bibfield{author}{\bibinfo{person}{Jessica Su}, \bibinfo{person}{Ansh Shukla}, \bibinfo{person}{Sharad Goel}, {and} \bibinfo{person}{Arvind Narayanan}.} \bibinfo{year}{2017}\natexlab{}.
\newblock \showarticletitle{De-Anonymizing Web Browsing Data with Social Networks}. \bibinfo{howpublished}{\url{https://doi.org/10.1145/3038912.3052714}}. In \bibinfo{booktitle}{\emph{Proceedings of the 26th International Conference on World Wide Web}} (Perth, Australia) \emph{(\bibinfo{series}{WWW '17})}. \bibinfo{publisher}{International World Wide Web Conferences Steering Committee}, \bibinfo{address}{Republic and Canton of Geneva, CHE}, \bibinfo{pages}{1261–1269}.
\newblock
\showISBNx{9781450349130}
\urldef\tempurl%
\url{https://doi.org/10.1145/3038912.3052714}
\showDOI{\tempurl}


\bibitem[Sweeney(2000)]%
        {sweeney2000simple}
\bibfield{author}{\bibinfo{person}{Latanya Sweeney}.} \bibinfo{year}{2000}\natexlab{}.
\newblock \showarticletitle{Simple demographics often identify people uniquely}.
\newblock \bibinfo{journal}{\emph{Health (San Francisco)}} \bibinfo{volume}{671}, \bibinfo{number}{2000} (\bibinfo{year}{2000}), \bibinfo{pages}{1--34}.
\newblock


\bibitem[Swichkow(2014)]%
        {prank}
\bibfield{author}{\bibinfo{person}{Brian Swichkow}.} \bibinfo{year}{2014}\natexlab{}.
\newblock \bibinfo{title}{{The Ultimate Retaliation: Pranking My Roommate With Targeted Facebook Ads}}.
\newblock \bibinfo{howpublished}{Ghost Influence}.
\newblock
\urldef\tempurl%
\url{http://ghostinfluence.com/the-ultimate-retaliation-pranking-my-roommate-with-targeted-facebook-ads/}
\showURL{%
\tempurl}
\newblock
\shownote{accessed on 29 January, 2023}.


\bibitem[Thepiratebay(2023)]%
        {piratebay}
\bibfield{author}{\bibinfo{person}{Thepiratebay}.} \bibinfo{year}{2023}\natexlab{}.
\newblock \bibinfo{booktitle}{}.
\newblock
\urldef\tempurl%
\url{thepiratebay.org}
\showURL{%
\tempurl}


\bibitem[Tim~Shipman(2018)]%
        {tim_shipman_2018}
\bibfield{author}{\bibinfo{person}{Political~Editor Tim~Shipman}.} \bibinfo{year}{2018}\natexlab{}.
\newblock \bibinfo{title}{{Labour HQ used Facebook ads to deceive Jeremy Corbyn}}.
\newblock \bibinfo{howpublished}{The Sunday Times}.
\newblock
\urldef\tempurl%
\url{https://www.thetimes.co.uk/article/labour-hq-used-facebook-ads-to-deceive-corbyn-3hvn0jzr8}
\showURL{%
\tempurl}
\newblock
\shownote{accessed on 29 January, 2023}.


\bibitem[Tricomi et~al\mbox{.}(2023)]%
        {Dota2AIA}
\bibfield{author}{\bibinfo{person}{Pier~Paolo Tricomi}, \bibinfo{person}{Lisa Facciolo}, \bibinfo{person}{Giovanni Apruzzese}, {and} \bibinfo{person}{Mauro Conti}.} \bibinfo{year}{2023}\natexlab{}.
\newblock \showarticletitle{Attribute Inference Attacks in Online Multiplayer Video Games: A Case Study on DOTA2}. In \bibinfo{booktitle}{\emph{Proceedings of the Thirteenth ACM Conference on Data and Application Security and Privacy}} (Charlotte, NC, USA) \emph{(\bibinfo{series}{CODASPY '23})}. \bibinfo{publisher}{Association for Computing Machinery}, \bibinfo{address}{New York, NY, USA}, \bibinfo{pages}{27–38}.
\newblock
\showISBNx{9798400700675}
\urldef\tempurl%
\url{https://doi.org/10.1145/3577923.3583653}
\showDOI{\tempurl}


\bibitem[Twitter({[n.\,d.]})]%
        {twittercustomaudiences}
\bibfield{author}{\bibinfo{person}{Twitter}.} \bibinfo{year}{[n.\,d.]}\natexlab{}.
\newblock \bibinfo{booktitle}{\emph{Intro to Custom Audiences}}.
\newblock


\bibitem[Venkatadri et~al\mbox{.}(2018)]%
        {venkatadriPIIfb}
\bibfield{author}{\bibinfo{person}{Giridhari Venkatadri}, \bibinfo{person}{Athanasios Andreou}, \bibinfo{person}{Yabing Liu}, \bibinfo{person}{Alan Mislove}, \bibinfo{person}{Krishna~P. Gummadi}, \bibinfo{person}{Patrick Loiseau}, {and} \bibinfo{person}{Oana Goga}.} \bibinfo{year}{2018}\natexlab{}.
\newblock \showarticletitle{Privacy Risks with Facebook's PII-Based Targeting: Auditing a Data Broker's Advertising Interface}. In \bibinfo{booktitle}{\emph{2018 IEEE Symposium on Security and Privacy (SP)}}. \bibinfo{pages}{89--107}.
\newblock
\urldef\tempurl%
\url{https://doi.org/10.1109/SP.2018.00014}
\showDOI{\tempurl}


\bibitem[Wheeler et~al\mbox{.}(2005)]%
        {pshyc_persuasion2}
\bibfield{author}{\bibinfo{person}{S. Wheeler}, \bibinfo{person}{Richard Petty}, {and} \bibinfo{person}{George Bizer}.} \bibinfo{year}{2005}\natexlab{}.
\newblock \showarticletitle{{Self-Schema Matching and Attitude Change: Situational and Dispositional Determinants of Message Elaboration}}.
\newblock \bibinfo{journal}{\emph{Journal of Consumer Research}}  \bibinfo{volume}{31} (\bibinfo{date}{March} \bibinfo{year}{2005}), \bibinfo{pages}{787--797}.
\newblock
\urldef\tempurl%
\url{https://doi.org/10.1086/426613}
\showDOI{\tempurl}


\bibitem[Zarras et~al\mbox{.}(2014)]%
        {zarrasmalvert}
\bibfield{author}{\bibinfo{person}{Apostolis Zarras}, \bibinfo{person}{Alexandros Kapravelos}, \bibinfo{person}{Gianluca Stringhini}, \bibinfo{person}{Thorsten Holz}, \bibinfo{person}{Christopher Kruegel}, {and} \bibinfo{person}{Giovanni Vigna}.} \bibinfo{year}{2014}\natexlab{}.
\newblock \showarticletitle{The Dark Alleys of Madison Avenue: Understanding Malicious Advertisements}. In \bibinfo{booktitle}{\emph{Proceedings of the 2014 Conference on Internet Measurement Conference}} (Vancouver, BC, Canada) \emph{(\bibinfo{series}{IMC '14})}. \bibinfo{publisher}{Association for Computing Machinery}, \bibinfo{address}{New York, NY, USA}, \bibinfo{pages}{373–380}.
\newblock
\showISBNx{9781450332132}
\urldef\tempurl%
\url{https://doi.org/10.1145/2663716.2663719}
\showDOI{\tempurl}


\end{thebibliography}

\appendix
\section*{Appendix}

\section{LinkedIn dashboard report for the proof of concept experiment ad campaigns}
\label{app:LinedIn_records}

\begin{figure*}[t]
    \centering
    \includegraphics[width=0.80\linewidth]{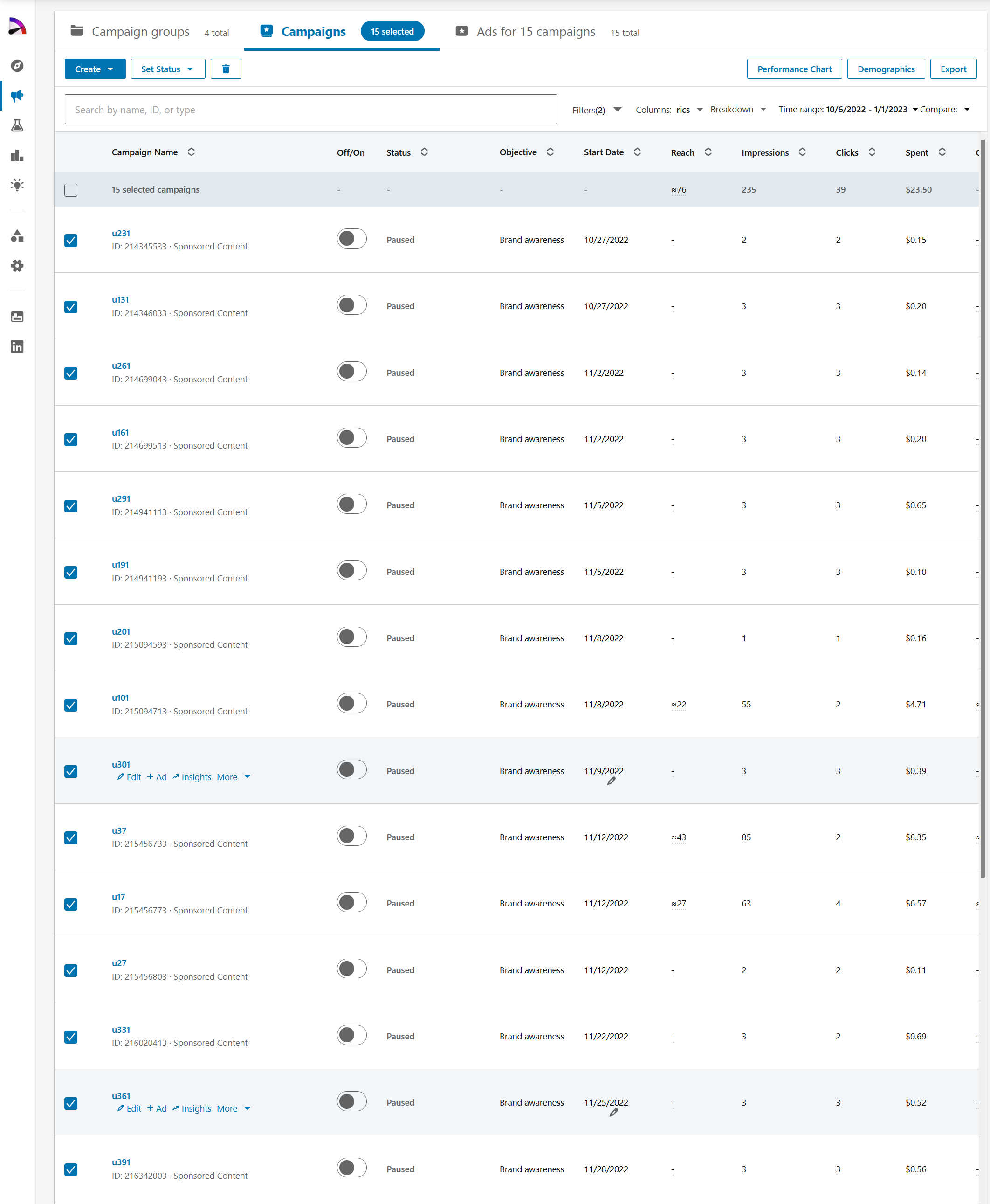}
    \caption{Screenshot of the LinkedIn Campaign Manager dashboard, including the results of the 15 ad campaigns executed in our proof of concept experiment.}
    \label{fig:ldrep}
    \Description{Screenshot showing the report of the LinkedIn Campaign Manager dashboard for the 15 ad campaigns executed in our proof of concept experiment.}
\end{figure*}

Some of the results reported in Table \ref{table:poc_results} were extracted from the LinkedIn Campaign Manager report for the different advertising campaigns delivered. Figure \ref{fig:ldrep} shows a snapshot of the information reported by LinkedIn for the 15 ad campaigns executed in our proof of concept experiment. It shows the ID of the campaign, the budget spent, the number of delivered ad impressions, and the number of clicks received. While we acknowledge that the report may not be 100\% accurate, the fact that impressions counted by the targeted individuals in our experiment always match the number of impressions reported by LinkedIn leads us to think that not only that the targeted users the only ones have received their corresponding ad, but also that the LinkedIn count is accurate, at least for audiences close to 1 user.

\section{Country distribution of users in our data sample}
\label{app:countries_breakdown}

Our dataset contains samples from  107 \color{black} different countries. Table \ref{table:samples_by_country} shows the breakdown of the number of users per country and dataset. About 49\% of the users in our dataset are from the United States. We acknowledge the fact that this circumstance in our data may lead to some biases in the results of our model and, therefore, to the estimation of $N$. However, the fact that the proof of concept experiment was targeting users in a different country than the US and the obtained results are aligned with the model outcome makes us confident that the potential bias (if any) may not be very relevant.  Also, Ds2 is not biased toward the United States, and the separate analysis of this dataset leads to similar results.

%Our intuition, based on the paper's results, is that our model serves as an upper bound for the number of skills needed to make a user unique. This may be because the United States is one of the countries contributing more LinkedIn users. Therefore, it seems reasonable to estimate that, in many cases, it will be easier to re-identify users reporting a different location than the US.

\section{Model fitting in the other considered scenarios and datasets}
\label{app:scenarios}

We apply our methodology to four different scenarios summarized in section \ref{subsec:scenarios} {and to each of the three datasets and the aggregation of them}.

\begin{figure*}[t]
\centering
	\begin{minipage}[t]{0.32\hsize}
		\centering
		\includegraphics[width=\columnwidth]{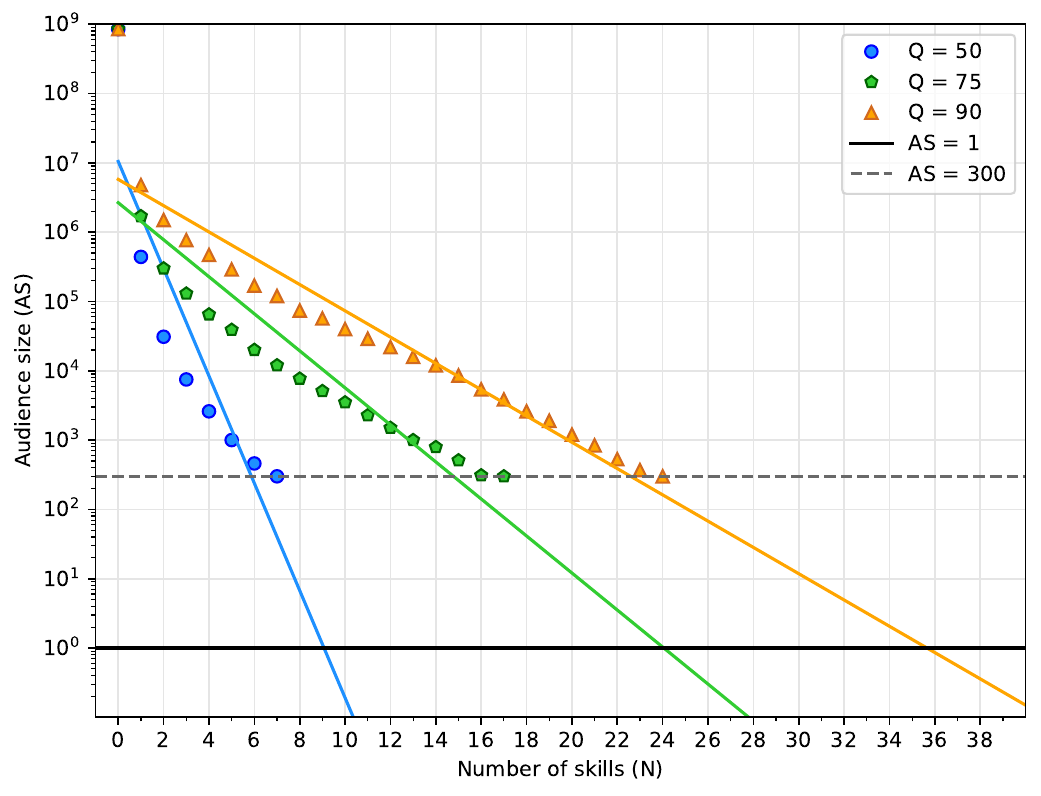}
            \caption{\blu Application of the methodology to the $Sk\_LP{\_Agg}$ scenario for $V\textsubscript{AS}(Q)$ with $Q = 50, 75$ and $90$.}
            \label{fig:sk_lp_line_fit_agg}
            \Description{Line graph showing the results of applying our methodology to our dataset, for the scenario in which we only use skills to define the audiences (and not the location), and we select the skills with the least popular approach (Sk_LP). The Y axis represents the audience size, while the X axis represents the number of skills used to define an audience. The points show the quantiles 50, 75, and 90 for each value in the X-axis, and the lines represent the fitting of the points for each quantile. Both the points and lines show how the audience size decreases as the audience is defined with more skills, and the lines estimate at what point the audience size would reach a value of one for each quantile.}
	\end{minipage}
    \hfill
	\begin{minipage}[t]{0.32\hsize}
		\includegraphics[width=\columnwidth]{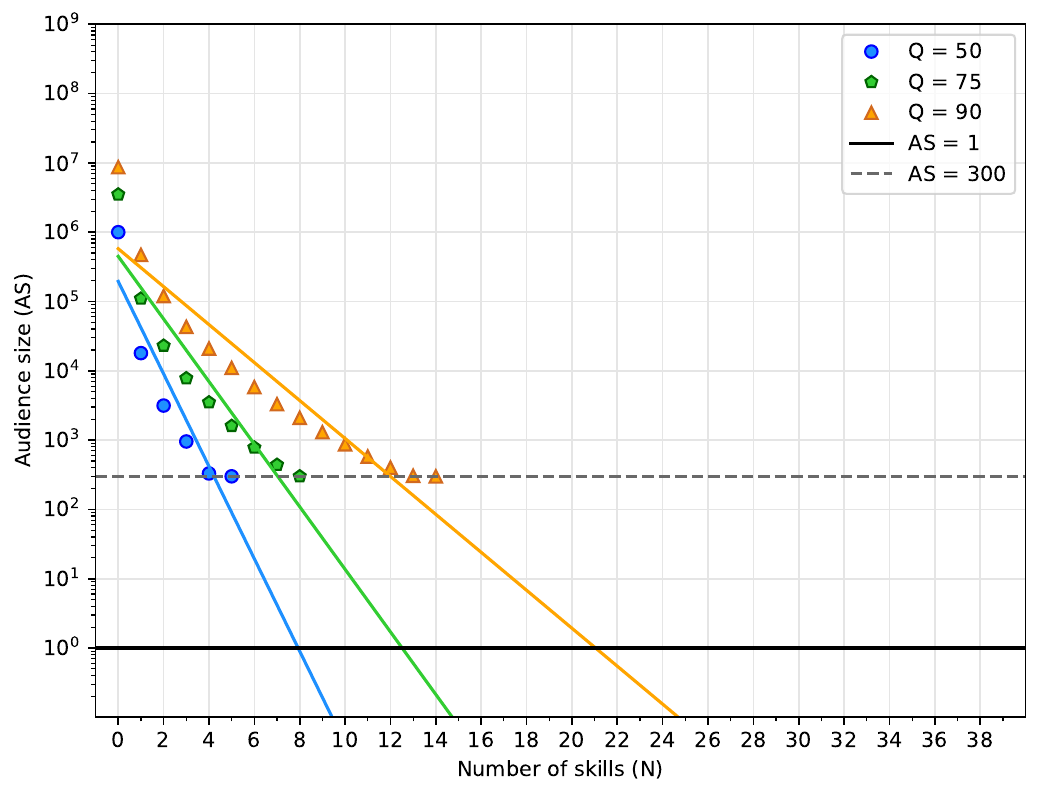}
            \caption{\blu Application of the methodology to the $Lo\_R{\_Agg}$ scenario for $V\textsubscript{AS}(Q)$ with $Q = 50, 75$ and $90$.}
            \label{fig:lo_r_line_fit_agg}
            \Description{Line graph showing the results of applying our methodology to our dataset, for the scenario in which we use the skills and the location to define the audiences, and we select the skills randomly (Lo_R). The Y axis represents the audience size, while the X axis represents the number of skills used to define an audience. The points show the quantiles 50, 75, and 90 for each value in the X-axis, and the lines represent the fitting of the points for each quantile. Both the points and lines show how the audience size decreases as the audience is defined with more skills, and the lines estimate at what point the audience size would reach a value of one for each quantile.}
    \end{minipage}
    \hfill
    \begin{minipage}[t]{0.32\hsize}
		\centering
            \includegraphics[width=\columnwidth]{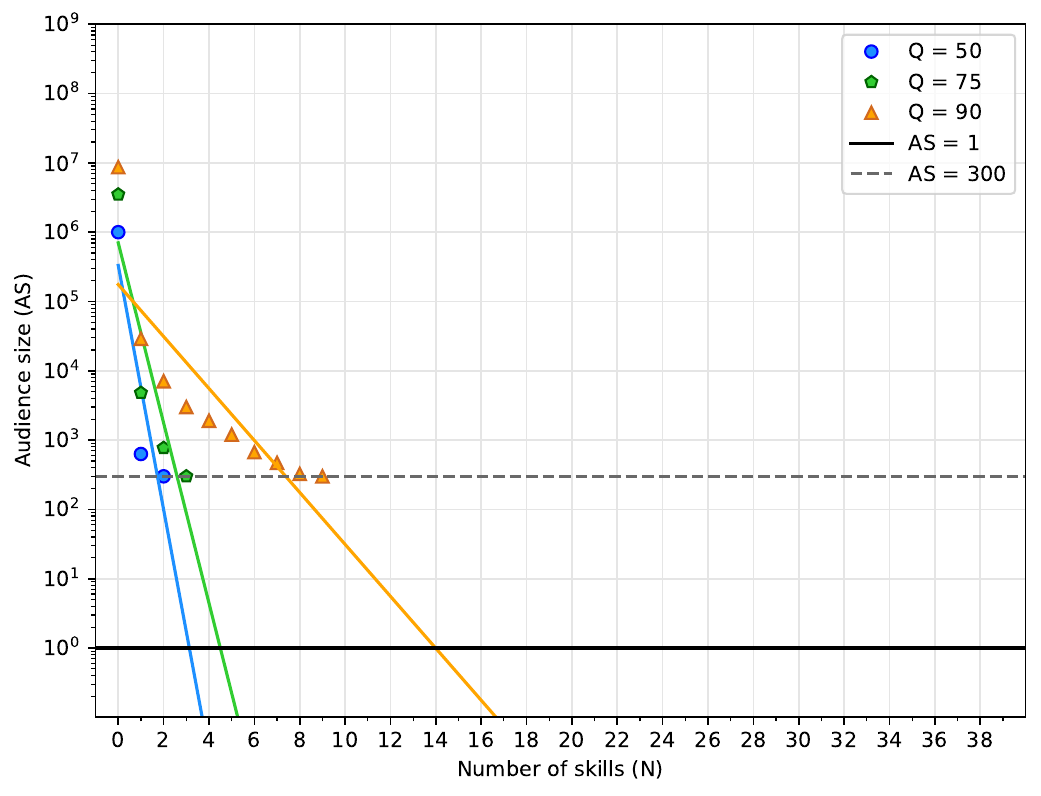}
            \caption{\blu Application of the methodology to the $Lo\_LP{\_Agg}$ scenario for $V\textsubscript{AS}(Q)$ with $Q = 50, 75$ and $90$.}
            \label{fig:lo_lp_line_fit_agg}
            \Description{Line graph showing the results of applying our methodology to our dataset, for the scenario in which we use the skills and the location to define the audiences, and we select the skills with the least popular approach (Lo_LP). The Y axis represents the audience size, while the X axis represents the number of skills used to define an audience. The points show the quantiles 50, 75, and 90 for each value in the X-axis, and the lines represent the fitting of the points for each quantile. Both the points and lines show how the audience size decreases as the audience is defined with more skills, and the lines estimate at what point the audience size would reach a value of one for each quantile.}
	\end{minipage}
\end{figure*}

This appendix shows the model fitting for the other three scenarios {and datasets} not shown in the paper body referred to as $(ii)$ $Sk\_LP\blu\_Agg$ (Figure \ref{fig:sk_lp_line_fit_agg}), we only use the least popular professional skills; $(iii)$ $Lo\_R\blu\_Agg$ (Figure \ref{fig:lo_r_line_fit_agg}), we use the location and professional skills selected at random; $(iv)$ $Lo\_LP\blu\_Agg$ (Figure \ref{fig:lo_lp_line_fit_agg}), we use the location and the least popular skills. {The other figures show the same scenarios but applied to each dataset separately.}  Table \ref{table:pessimistic} shows the $R^2$ values for all the line fittings.

%{We also acknowledge that for the $P=90$, for the $Lo\_{LP}$ we observe a rather large difference of 9 skills between Ds1 ($N_{90}=18.2$) and Ds3 ($N_{90}=8.9$). Thinking about the way we collect each dataset, we think that Ds1 may contain more similar users, as the BFS search provides related users. If some skills are shared by more of the users in the sample, this can affect the results, especially for the cases where the skills are selected by popularity, where the more popular skills will tend to be placed in the last positions, and if those skills are the same for several users, the samples will report similar AS, biasing the quantiles (especially higher ones, where the more popular skills come into play for the $N_P$ estimation) to that audience size corresponding to that similar, biased users.}

\begin{table*}[t]
\small
\begin{tabular}{l r r r r l r r r r}
\toprule
\textbf{Country} & \textbf{Ds1} & \textbf{Ds2} & \textbf{Ds3} & \textbf{DsAgg} & \textbf{Country} & \textbf{ Ds1} & \textbf{Ds2} & \textbf{Ds3} & \textbf{DsAgg}\\
\midrule

United States & 1282 & 29 &  334 &  1645 &  Norway &  0 &  9 &  0 &  9\\

Canada & 92 &  17 &  33 &  142 & Uruguay & 1 &  8 &  0 &  9\\

United Kingdom & 60 &  21 &  39 &  120 & Ireland & 3 &  3 &  2 &  8 \\

India & 24 &  32 &  40 &  96  & New Zeland &  0 &  8 &  0 &  8 \\

Spain & 40 &  22 &  33 &  95  & Costa Rica &  0 &  8 &  0 &  8 \\

Australia & 10 &  40 &  17 &  67  & Croatia &  0 &  5 &  3 &  8 \\

Italy & 19 &  20 &  10 &  49 & Malaysia & 2 &  5 &  1 &  8\\

United Arab Emirates & 9 &  21 &  12 &  42 & Hungary & 1 &  6 &  1 &  8\\

Germany & 7 &  21 &  5 &  33 &  Kazakhstan &  0 &  8 &  1 &  8\\

Brazil & 1 &  31 &  0 &  32  & Guatemala &  0 &  7 &  0 &  7 \\

Netherlands & 6 &  22 &  2 &  30 & Taiwan & 2 &  5 &  0 &  7\\

Indonesia & 3 &  24 &  3 &  30 & Thailand & 1 &  6 &  0 &  7\\

Portugal & 2 &  27 &  1 &  30 & Ghana & 1 &  6 &  0 &  7\\

Belgium & 9 &  14 &  4 &  27 &  Tunisa &  0 &  7 &  0 &  7\\

France & 7 &  17 &  3 &  27 & Ethiopia & 5 &  1 &  0 &  6 \\

Pakistan & 2 &  17 &  8 &  27 &  Sri Lanka &  0 &  6 &  0 &  6 \\

Turkey & 5 &  18 &  3 &  26 &  Honduras &  0 &  6 &  0 &  6 \\

Israel & 13 &  9 &  3 &  25 &  Serbia &  0 &  6 &  0 &  6 \\

Greece &  11 &  13 &  0 &  24 &  Lebanon &  0 &  5 &  1 &  6\\

 Bulgaria &  0 &  22 &  2 &  24 &  Slovakia &  0 &  6 &  0 &  6\\

Mexico & 5 &  12 &  5 &  22 &  Angola &  0 &  6 &  0 &  6\\

Panama & 5 &  17 &  0 &  22 & Uganda & 4 &  1 &  0 &  5 \\

Switzerland & 2 &  18 &  2 &  22 &  Puerto Rico &  0 &  4 &  1 &  5 \\

South Africa & 1 &  8 &  12 &  21 & Kuwait & 2 &  2 &  1 &  5\\

Chile & 8 &  12 &  0 &  20 &  Zimbabwe &  0 &  3 &  2 &  5\\

 Saudi Arabia &  0 &  15 &  5 &  20 &  El Salvador &  0 &  4 &  0 &  4 \\

Venezuela & 4 &  15 &  0 &  19 & Trinidad and Tobago & 1 &  2 &  1 &  4\\

 Peru &  0 &  16 &  2 &  18 & Bolivia & 1 &  3 &  0 &  4\\

Denmark & 1 &  14 &  2 &  17 &  South Korea &  0 &  4 &  0 &  4\\

 Sweden &  0 &  12 &  4 &  16 &  Vietnam &  0 &  3 &  1 &  4\\

Poland & 1 &  15 &  0 &  16 & Slovenia & 3 &  0 &  0 &  3\\

Colombia & 2 &  13 &  0 &  15 &  Nicaragua &  0 &  3 &  0 &  3 \\

 Jordan &  0 &  15 &  0 &  15 &  Tanzania &  0 &  3 &  0 &  3 \\

Finland & 1 &  9 &  5 &  15 &  Nepal &  0 &  3 &  0 &  3 \\

Philippines & 5 &  6 &  3 &  14 &  Senegal &  0 &  2 &  1 &  3 \\

Romania & 4 &  5 &  5 &  14 &  Paraguay &  0 &  3 &  0 &  3\\

 Singapore &  0 &  14 &  0 &  14 &  Cameroon &  0 &  2 &  1 &  3\\

 Bangladesh &  0 &  12 &  2 &  14 &  Palestinian Authority &  0 &  0 &  2 &  2 \\

Japan & 3 &  10 &  0 &  13 & Cyprus & 2 &  0 &  0 &  2\\

 Qatar &  0 &  13 &  0 &  13 & Bosnia and Herzegovina & 1 &  0 &  1 &  2\\

Argentina & 3 &  8 &  1 &  12 & Russia & 2 &  0 &  0 &  2\\

Kenya & 1 &  10 &  1 &  12 &  Afghanistan &  0 &  1 &  1 &  2 \\

 Ecuador &  0 &  11 &  0 &  11 &  Oman &  0 &  2 &  0 &  2\\

 Czechia &  0 &  9 &  2 &  11 &  Monaco &  0 &  0 &  2 &  2\\

China & 1 &  8 &  2 &  11 &  Cambodia &  0 &  1 &  0 &  1\\

 Dominican Republic &  0 &  11 &  0 &  11 &  Zambia &  0 &  1 &  0 &  1 \\

Algeria & 1 &  10 &  0 &  11 &  Mauritius &  0 &  0 &  1 &  1 \\

 Egypt &  0 &  7 &  4 &  11 &  Jamaica &  0 &  0 &  1 &  1 \\

Nigeria & 1 &  8 &  2 &  11 & Malta & 1 &  0 &  0 &  1\\

 Iraq &  0 &  10 &  0 &  10 &  Bhutan &  0 &  0 &  1 &  1\\

Austria & 1 &  9 &  0 &  10 & Luxembourg & 1 &  0 &  0 &  1\\

Ukraine & 1 &  8 &  1 &  10 &  Belarus &  0 &  1 &  0 &  1\\

 Morocco &  0 &  9 &  1 &  10 &  Georgia &  0 &  1 &  0 &  1\\

Lithuania & 3 &  2 &  4 &  9 & \textbf{Not labeled} & \textbf{9} &  \textbf{18} &  \textbf{8} &  \textbf{35}\\

& & & & & \textbf{Total} & \textbf{1699} &  \textbf{1002} &  \textbf{651} &  \textbf{3352}\\

\bottomrule
\end{tabular}
\caption{Distribution of users in our datasets \color{black}per country.}
\label{table:samples_by_country}
\Description{Table that shows the breakdown of our dataset by country (according to the reported location in each profile).}
\end{table*}

\begin{figure*}[t]
\centering
        \begin{minipage}[t]{0.32\hsize}
		\centering
		\includegraphics[width=\columnwidth]{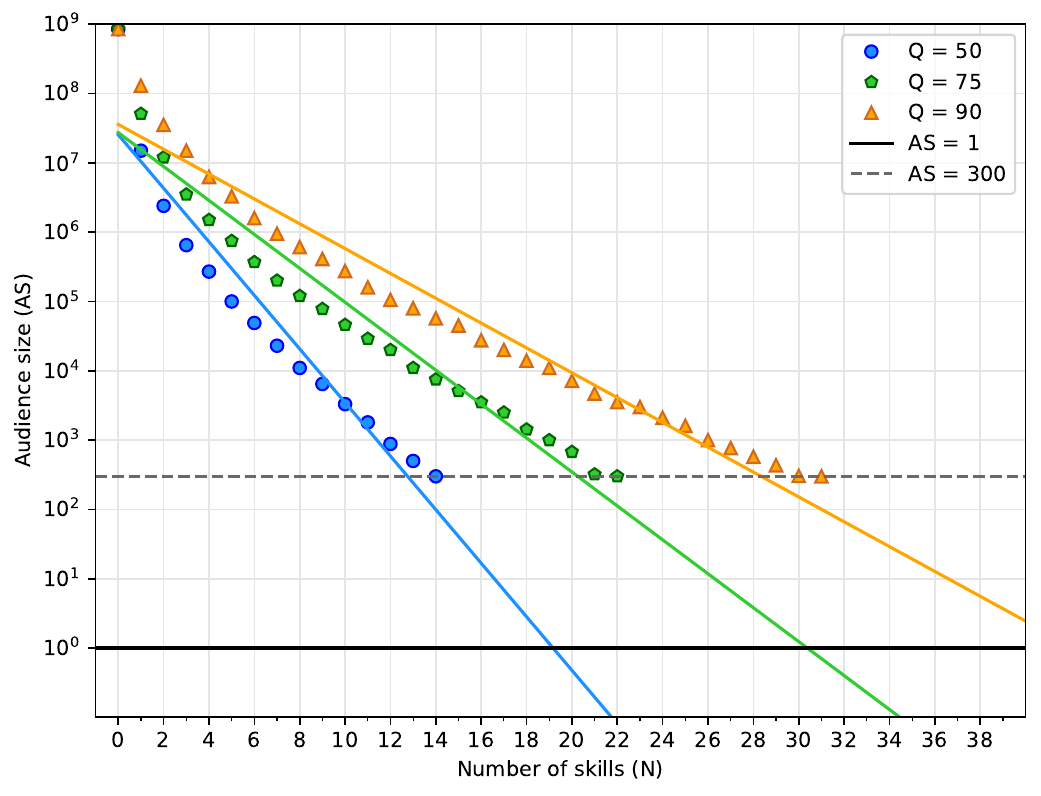}
        \caption{\blu Application of the methodology to the $Sk\_R{\_Ds1}$ scenario for $V\textsubscript{AS}(Q)$ with $Q = 50, 75$ and $90$.}
        \label{fig:sk_r_line_fit_ds1}
        \Description{}
	\end{minipage}
	\begin{minipage}[t]{0.32\hsize}
		\centering
		\includegraphics[width=\columnwidth]{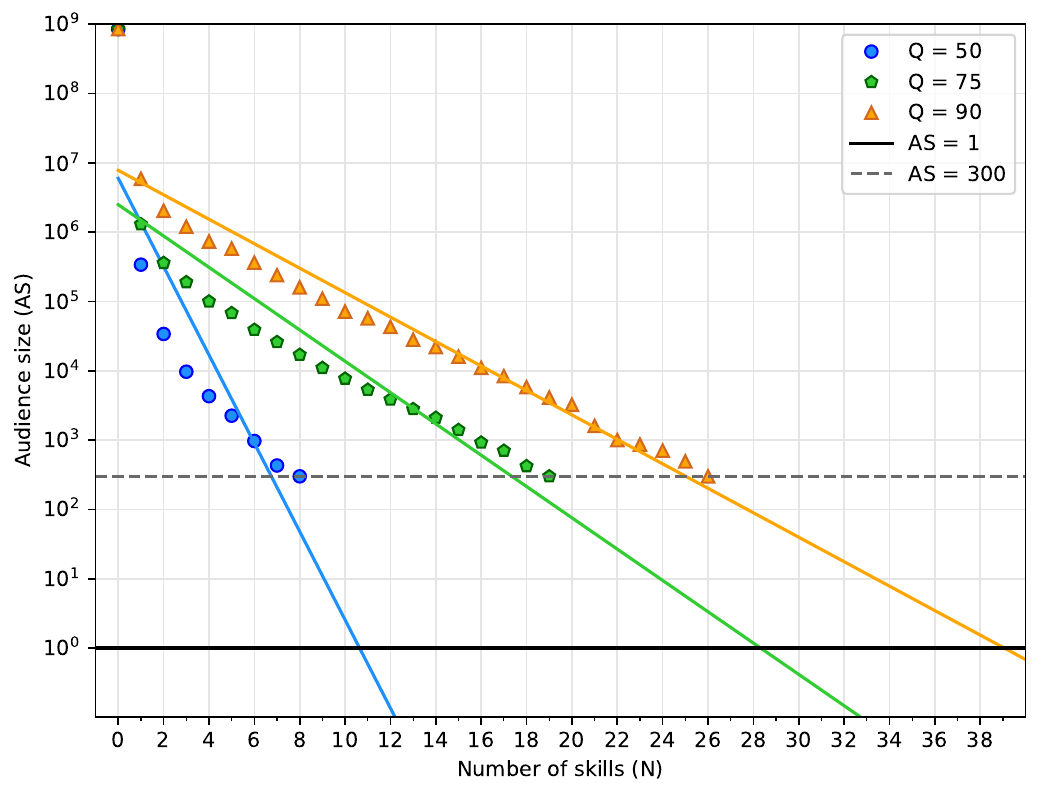}
            \caption{\blu Application of the methodology to the $Sk\_LP{\_Ds1}$ scenario for $V\textsubscript{AS}(Q)$ with $Q = 50, 75$ and $90$.}
            \label{fig:sk_lp_line_fit_ds1}
            \Description{}
	\end{minipage}
        \hfill
	\begin{minipage}[t]{0.32\hsize}
		\includegraphics[width=\columnwidth]{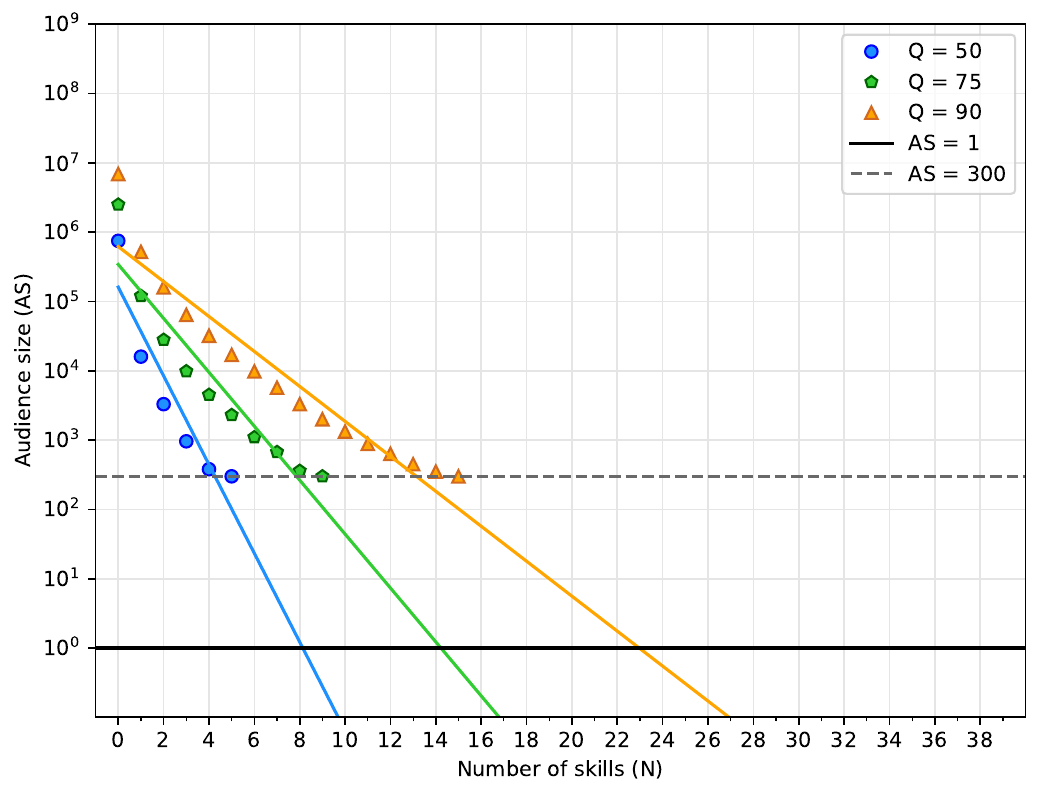}
            \caption{Application of the methodology to the $Lo\_R{\_Ds1}$ scenario for $V\textsubscript{AS}(Q)$ with $Q = 50, 75$ and $90$.}
            \label{fig:lo_r_line_fit_ds1}
            \Description{}
    \end{minipage}
    \hfill
\end{figure*}

\begin{figure*}[t]
\centering
	\begin{minipage}[t]{0.32\hsize}
		\centering
        \includegraphics[width=\columnwidth]{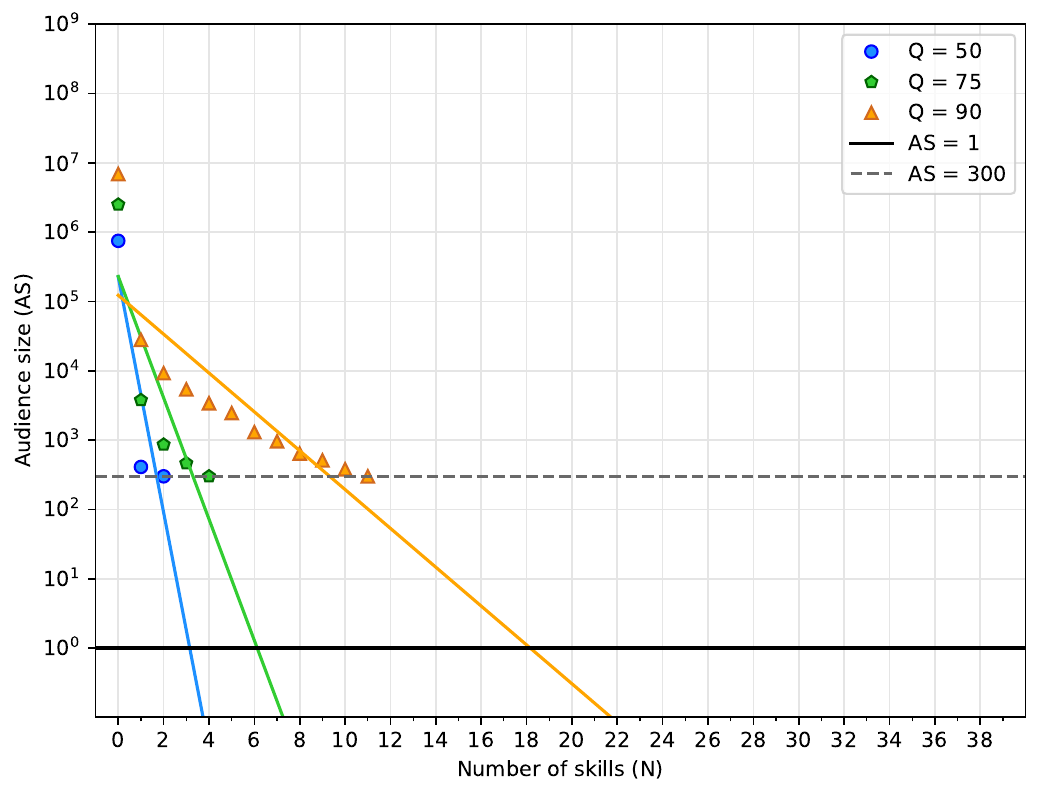}
        \caption{\blu Application of the methodology to the $Lo\_LP{\_Ds1}$ scenario for $V\textsubscript{AS}(Q)$ with $Q = 50, 75$ and $90$.}
        \label{fig:lo_lp_line_fit_ds1}
        \Description{}
	\end{minipage}
        \hfill
        \begin{minipage}[t]{0.32\hsize}
		\centering
		\includegraphics[width=\columnwidth]{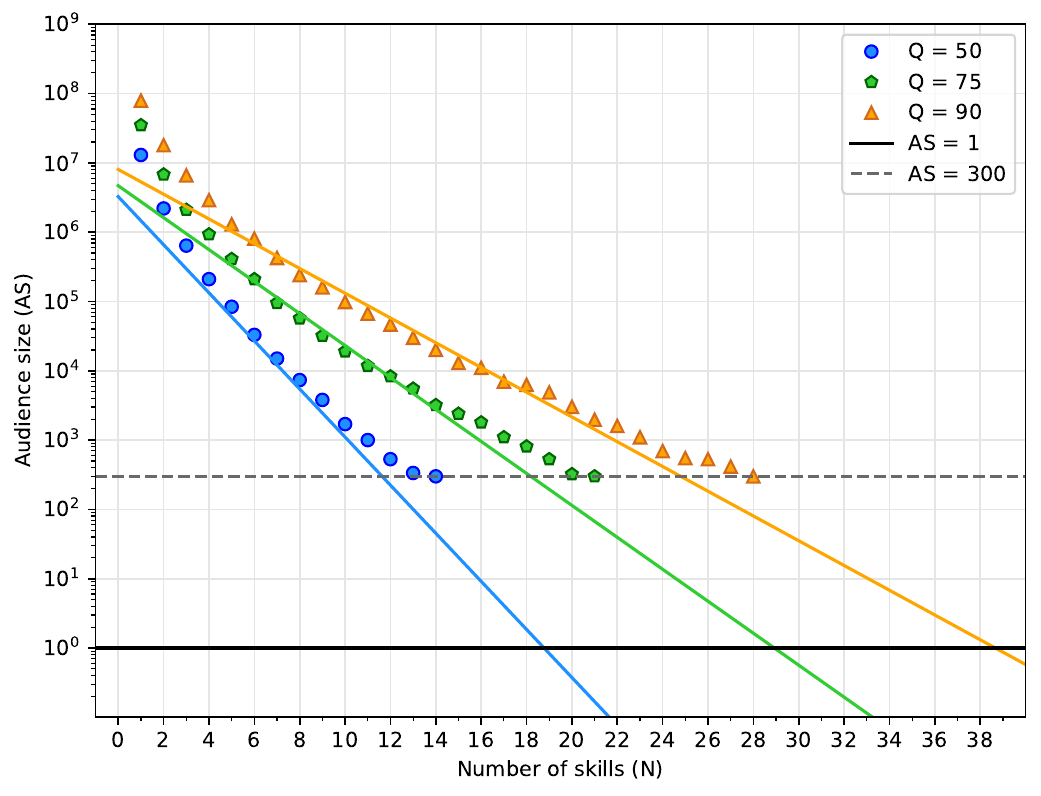}
        \caption{\blu Application of the methodology to the $Sk\_R{\_Ds2}$ scenario for $V\textsubscript{AS}(Q)$ with $Q = 50, 75$ and $90$.}
        \label{fig:sk_r_line_fit_ds2}
        \Description{}
	\end{minipage}
        \hfill
        \begin{minipage}[t]{0.32\hsize}
		\centering
		\includegraphics[width=\columnwidth]{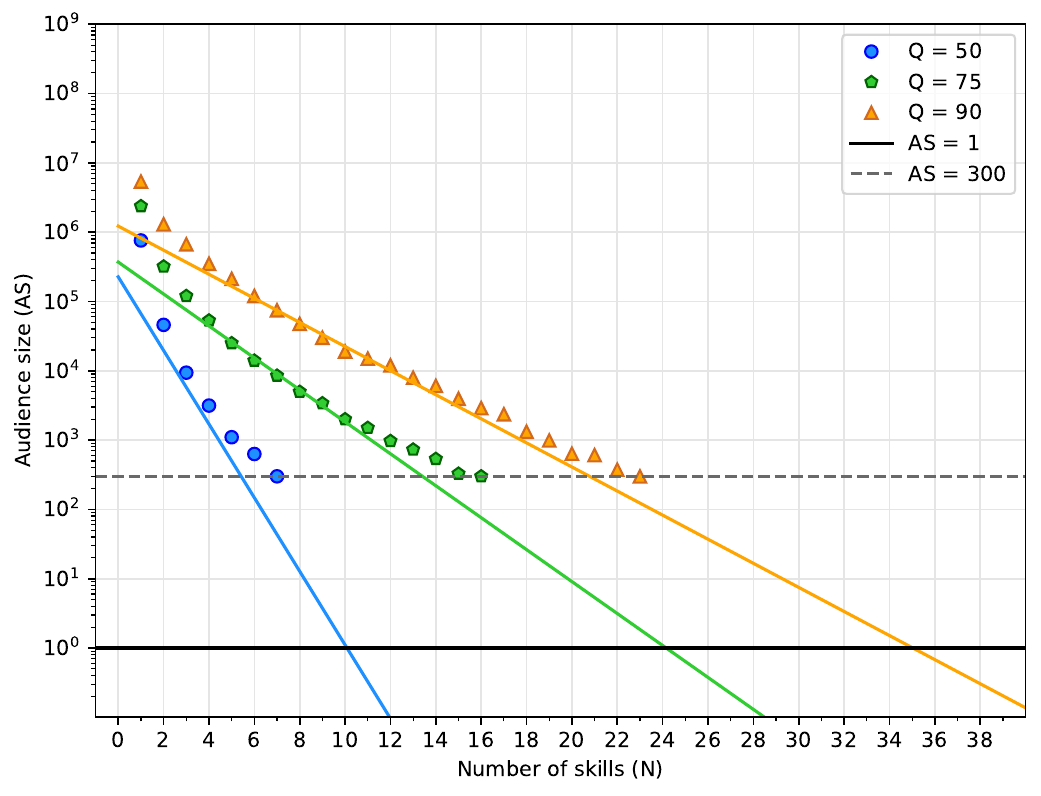}
        \caption{\blu Application of the methodology to the $Sk\_LP{\_Ds2}$ scenario for $V\textsubscript{AS}(Q)$ with $Q = 50, 75$ and $90$.}
        \label{fig:sk_lp_line_fit_ds2}
        \Description{}
	\end{minipage}
 
\end{figure*}

\begin{figure*}[t]
\centering
	\begin{minipage}[t]{0.32\hsize}
		\centering
        \includegraphics[width=\columnwidth]{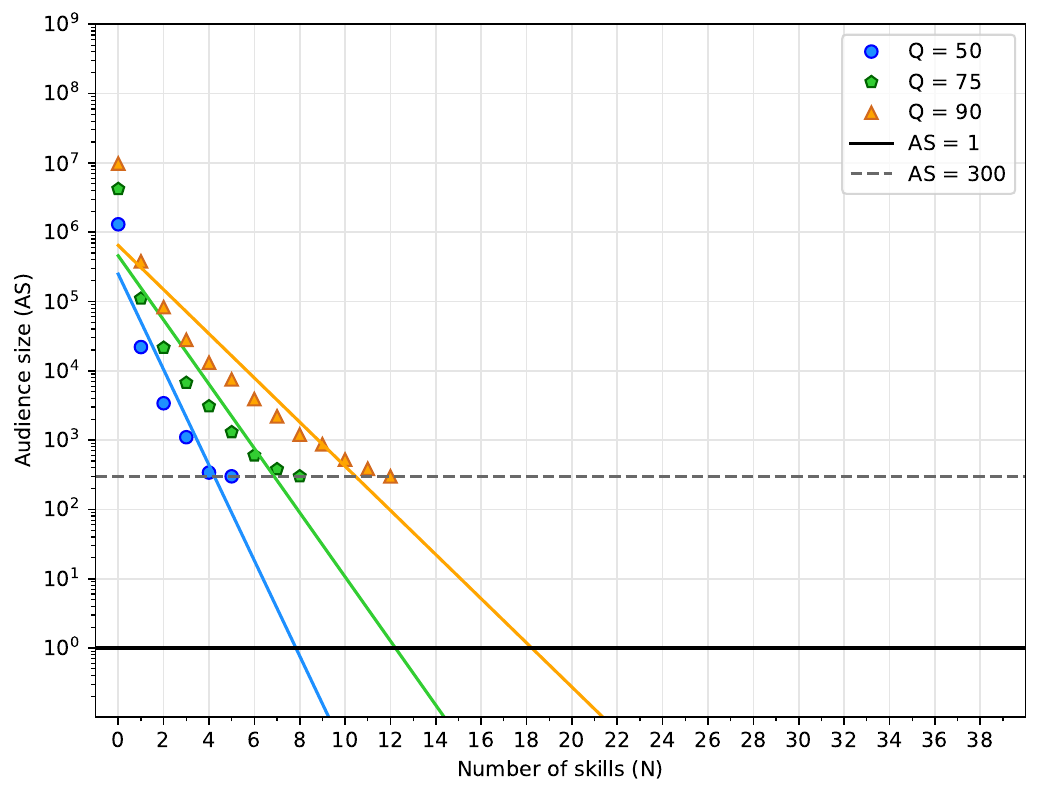}
        \caption{\blu Application of the methodology to the $Lo\_R{\_Ds2}$ scenario for $V\textsubscript{AS}(Q)$ with $Q = 50, 75$ and $90$.}
        \label{fig:lo_r_line_fit_ds2}
        \Description{}
	\end{minipage}
        \hfill
        \begin{minipage}[t]{0.32\hsize}
		\centering
		\includegraphics[width=\columnwidth]{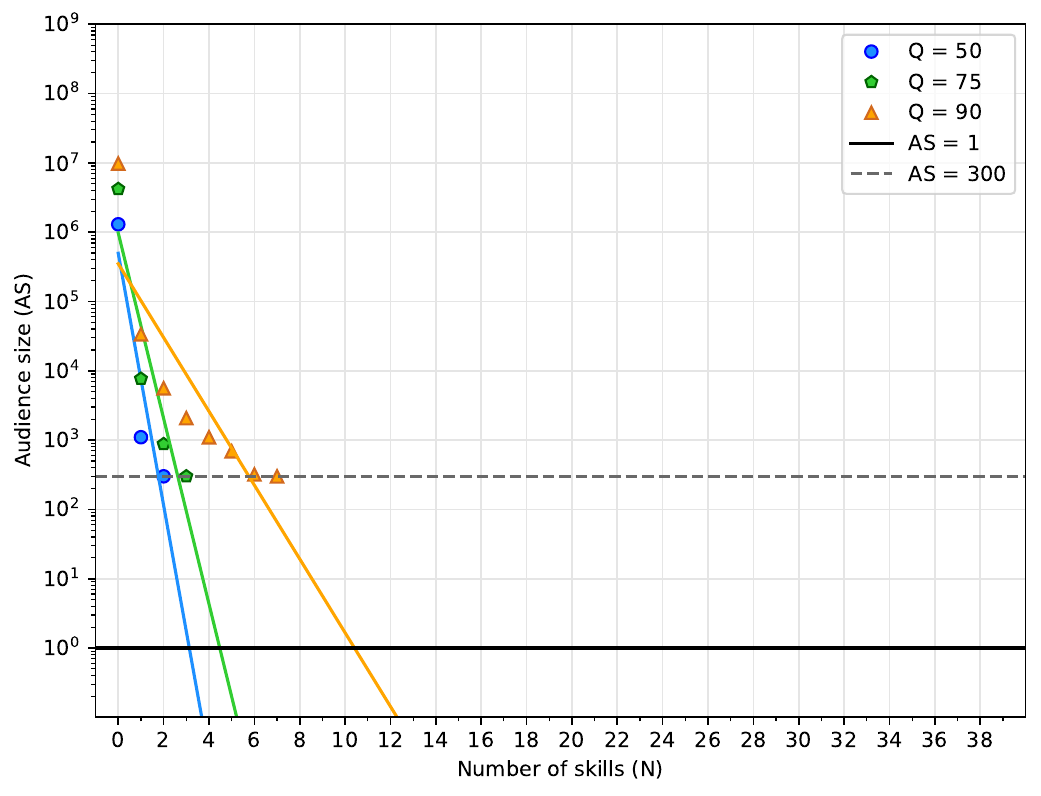}
        \caption{\blu Application of the methodology to the $Lo\_LP{\_Ds2}$ scenario for $V\textsubscript{AS}(Q)$ with $Q = 50, 75$ and $90$.}
        \label{fig:lo_lp_line_fit_ds2}
        \Description{}
	\end{minipage}
        \hfill
        \begin{minipage}[t]{0.32\hsize}
		\centering
		\includegraphics[width=\columnwidth]{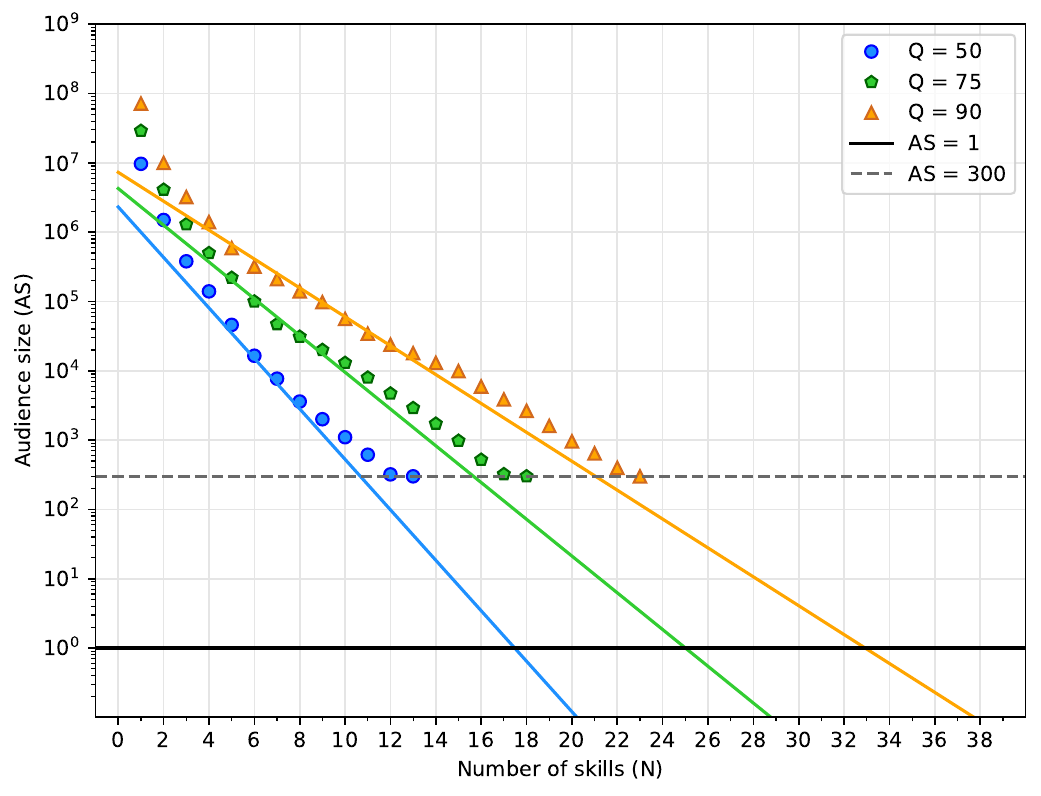}
        \caption{\blu Application of the methodology to the $Sk\_R{\_Ds3}$ scenario for $V\textsubscript{AS}(Q)$ with $Q = 50, 75$ and $90$.}
        \label{fig:sk_r_line_fit_ds3}
        \Description{}
	\end{minipage}
 
\end{figure*}

\begin{figure*}[t]
\centering
	\begin{minipage}[t]{0.32\hsize}
		\centering
		\includegraphics[width=\columnwidth]{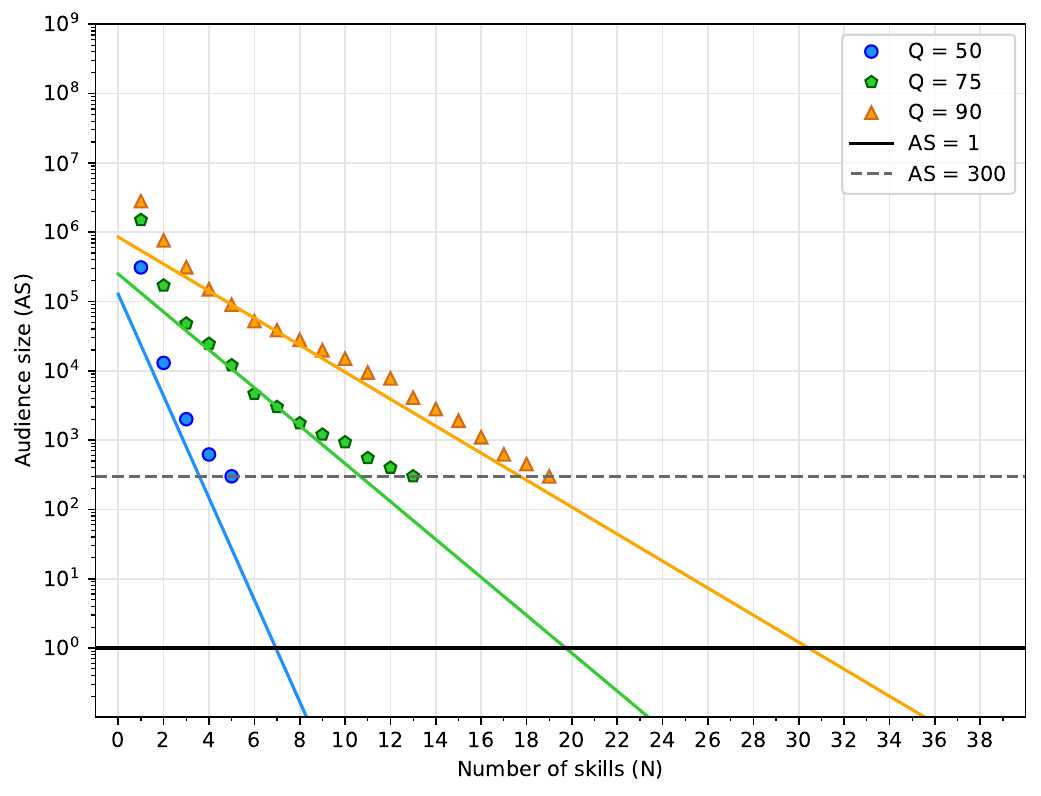}
            \caption{\blu Application of the methodology to the $Sk\_LP{\_Ds3}$ scenario for $V\textsubscript{AS}(Q)$ with $Q = 50, 75$ and $90$.}
            \label{fig:sk_lp_line_fit_ds3}
            \Description{}
	\end{minipage}
    \hfill
	\begin{minipage}[t]{0.32\hsize}
		\includegraphics[width=\columnwidth]{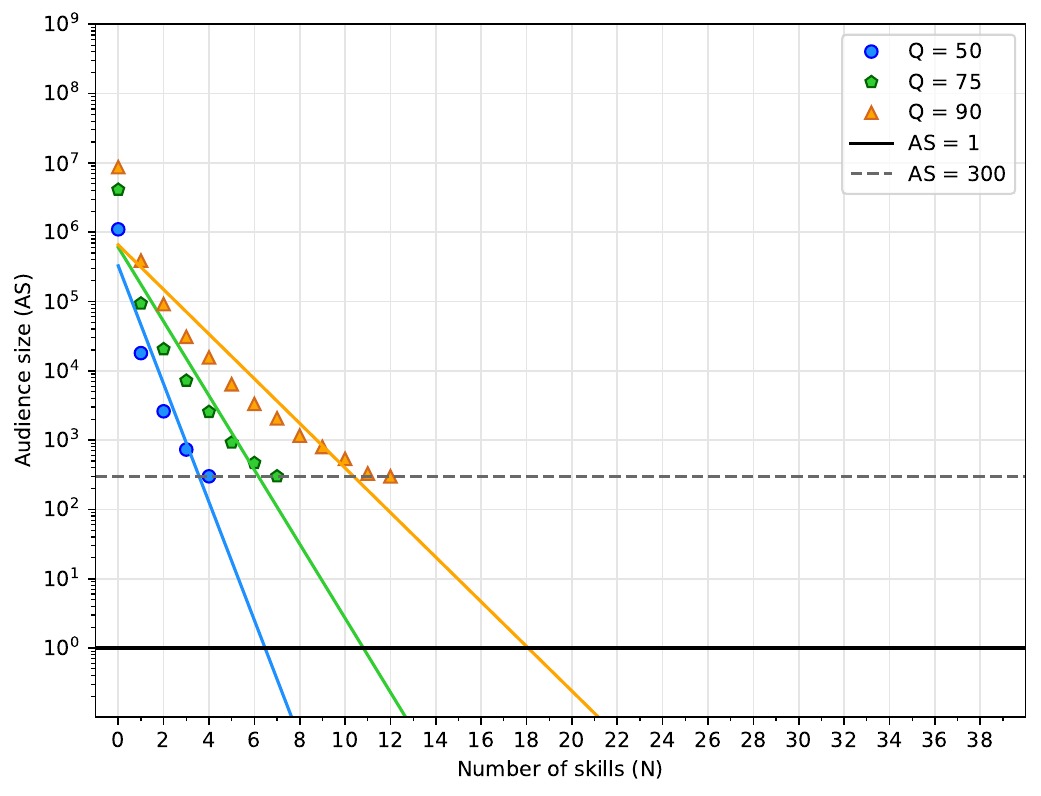}
            \caption{\blu Application of the methodology to the $Lo\_R{\_Ds3}$ scenario for $V\textsubscript{AS}(Q)$ with $Q = 50, 75$ and $90$. }
            \label{fig:lo_r_line_fit_ds3}
            \Description{}
    \end{minipage}
    \hfill
    \begin{minipage}[t]{0.32\hsize}
		\centering
            \includegraphics[width=\columnwidth]{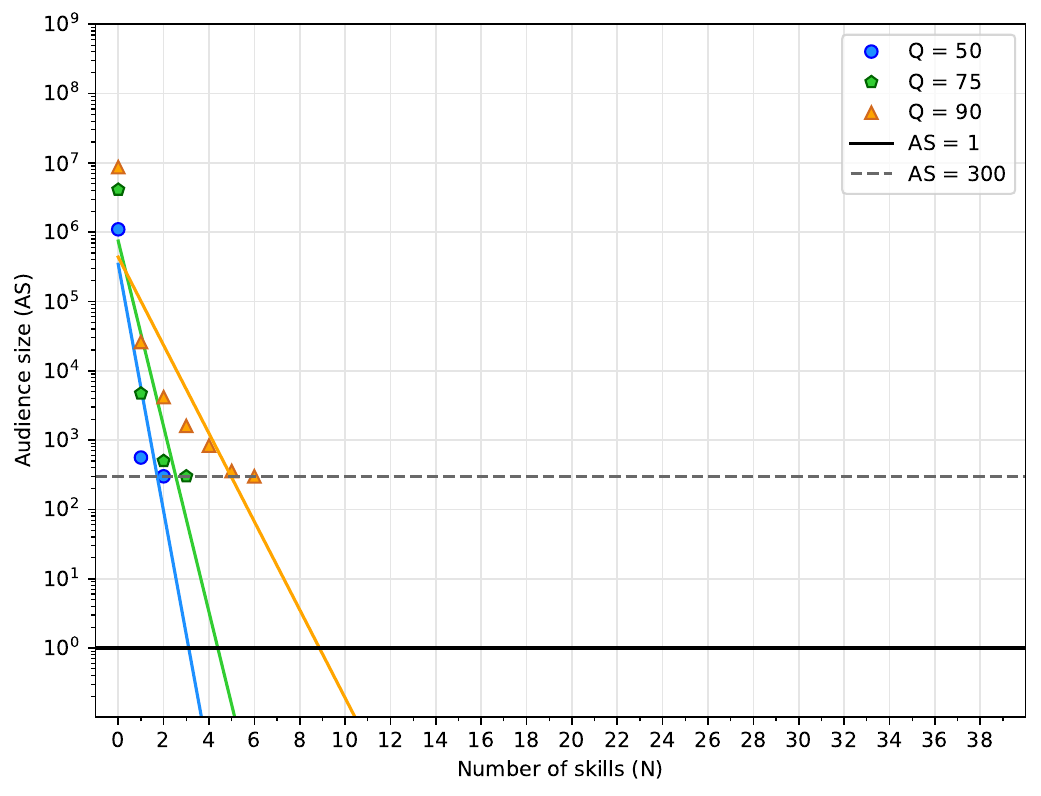}
            \caption{\blu Application of the methodology to the $Lo\_LP{\_Ds3}$ scenario for $V\textsubscript{AS}(Q)$ with $Q = 50, 75$ and $90$.}
            \label{fig:lo_lp_line_fit_ds3}
            \Description{}
	\end{minipage}
\end{figure*}

\end{document}